\documentclass[onecolumn,bibyear]{aa}
\usepackage[varg]{txfonts}

\usepackage{natbib}
\usepackage{float}
\usepackage{graphicx}
\usepackage{color,amssymb}
\usepackage{hyperref}
\usepackage{url}
\usepackage{ae,aecompl}
\newcommand{\cm}{cm$^{-1}$}

\begin{document}
\title{The ExoMolOP Database: Cross-sections and k-tables for Molecules of Interest in High-Temperature Exoplanet Atmospheres}

\titlerunning{ExoMolOP: Cross-section and k-table database}
	
\author{Katy L. Chubb\inst{1,2}\thanks{email: k.l.chubb@sron.nl, katy.chubb.14@ucl.ac.uk} \and Marco Rocchetto\inst{2} \and Sergei N. Yurchenko\inst{2}  \and Michiel Min\inst{1} \and Ingo Waldmann\inst{2} \and Joanna K. Barstow\inst{2} \and Paul Molli\`ere\inst{3} \and Ahmed F. Al-Refaie\inst{2} \and Mark W. Phillips\inst{4}  \and Jonathan Tennyson\inst{2}}

\institute{SRON Netherlands Institute for Space Research, Sorbonnelaan 2, 3584 CA, Utrecht, Netherlands \and Department of Physics and Astronomy, University College London, London, WC1E 6BT, UK \and Max-Planck-Institut f{\"u}r Astronomie , K{\"o}nigstuhl 17, 69117 Heidelberg, Germany \and Astrophysics Group, School of Physics, University of Exeter, Stocker Road, Exeter, EX4 4QL, UK} 

\abstract{A publicly available database of opacities for molecules of astrophysical interest, ExoMolOP, has been compiled for over 80 species, based on the latest line list data from the ExoMol, HITEMP and MoLLIST databases. These data are generally suitable for characterising high temperature exoplanet or cool stellar/substellar atmospheres, and have been computed at a variety of pressures and temperatures, with a few molecules included at room-temperature only from the HITRAN database. The data are formatted in different ways for four different exoplanet atmosphere retrieval codes; ARCiS, TauREx, NEMESIS and petitRADTRANS, and include both cross-sections (at R~=~$\frac{\lambda}{\Delta \lambda}$~=~15,000) and k-tables (at R~=~$\frac{\lambda}{\Delta \lambda}$~=~1000) for the 0.3~-~50$\mu$m wavelength region. Opacity files can be downloaded and used directly for these codes. Atomic data for alkali metals Na and K are also included, using data from the NIST database and the latest line shapes for the resonance lines. Broadening parameters have been taken from the literature where available, or from those for a known molecule with similar molecular properties where no broadening data are available. The data are available from www.exomol.com.}


\maketitle

\section{Introduction}\label{sec:intro}

There are now a large number of molecular line list data available for characterising hot exoplanet or cool stellar/substellar atmospheres, largely due to databases such as ExoMol~\citep{jt528,jt631}, HITEMP~\citep{HITEMP,jt763}, MoLLIST~\citep{MOLLIST}, and TheoReTs~\citep{TheoReTS}. Line lists are independent of temperature and pressure, and so provide the most efficient way of storing the information required for characterising high-temperature astrophysical atmospheres. In order to convert this line list data into opacities (cross-sections or k-tables), software such as ExoCross \citep{ExoCross}, is required to convert from a pressure and temperature independent line list to cross-section data at a particular pressure and temperature. If a large number of pressures and temperatures are required for a large number of molecules this can be  a computationally demanding task. The present opacity database was formed in order to help reduce the computational effort of the community and to allow quick download and use of the data for many molecules designed specifically for use in atmospheric retrieval codes. The data are stored in formats which are exactly compatible for use with four different retrieval codes; ARCiS~\citep{20MiOrCh.arcis}, TauREx~\citep{15WaTiRo.taurex,15WaRoTi.taurex,19AlChWa.taurex}, NEMESIS~\citep{NEMESIS}, and petitRADTRANS~\citep{19MoWaBo.petitRADTRANS}. Retrieval codes such as these have their own processes for the computation of opacities, but have previously been limited to a sub-section of the molecules for which there is data available. The data format required to input the opacities into each retrieval code is detailed in Section~\ref{sec:ret_codes}, with the intention that the data files are sufficiently easy to manipulate and reformat for use in any general retrieval code. There are tools available online, such as the exo-k library\footnote{\url{http://perso.astrophy.u-bordeaux.fr/~jleconte/exo_k-doc/index.html}} (Leconte et al., in prep) which enable conversion between different formats, some of which are those used in this work. Many other works have computed opacities for use in radiative transfer retrieval and atmospheric modelling codes; see, for example, \cite{09ShFoLi.exo,08FrMaLo.exo,14FrLuFo.exo,19LeTaGr.exo,14AmBaTr.exo,17KeLuOw.exo,15GrHexx,19MaKiMe.exo,20KiHeOr.exo,12AlHoFr.exo,07ShBu.exo,13LiWoZh.exo,17GaMa.exo,20PhTrBa.exo,SCANonline,KURonline}. 





The database of cross-sections and k-tables presented in this work were utilised in a recent study by \cite{20ChMiKa.wasp43b} re-examining the transmission spectra of ``Hot Jupiter'' exoplanet WASP-43b. It was found that AlO, which had not previously been considered in similar analyses of the transmission spectra~\citep{14KrBeDe.wasp43b,17StLiBe.wasp43b,18FiHe.exo,18TsWaZi.exo,19WeLoMe.wasp43b,19IrPaTa.wasp43b}, was the molecule which fitted the data to the highest level of confidence
 out of all molecules for which high-temperature opacity data currently exists in the infra-red region covered by the HST WFC3 instrument~\citep{13Bean.hst}. Other molecules with absorption features in this 1.1~-~1.7~$\mu$m region include C$_2$H$_2$, HCN, FeH, NH$_3$, ScH, VO, and TiH~\citep{18TeYu.exo}. Opacities from this database have also been used in works related to the investigation of ions in the thermospheres of planets \citep{20BoCaCh.exo}, the modelling of Brown Dwarf atmospheres \citep{20LeCaCh.exo}, and other works related to exoplanet atmospheres \citep{20Tayloretal,20MiOrCh.arcis}. 


This paper is structured as follows. Section~\ref{sec:linelists} gives an introduction to the line lists used in the present database and their data format. Section~\ref{sec:opacities} discusses the computation of cross-sections and k-tables from these line lists, including details of line broadening parameters used for each species in Section~\ref{sec:broad}. Section~\ref{sec:ret_codes} gives an overview of the four retrieval codes for which these opacities are formatted, and the data format specifications of each opacity file. The line lists sources used for each species and their properties are detailed in Section~\ref{sec:sources}, with comments in Section~\ref{sec:comments_tables}, and isotopologues in Section~\ref{sec:isos}. The wavelength coverage of the database is briefly discussed in Section~\ref{sec:vis_UV}. Section~\ref{sec:exomolOP} summarises the ExoMolOP database, including access and upkeep. The opacity requirements for high-resolution studies are discussed in Section~\ref{sec:highres}. We give our conclusions in Section~\ref{sec:conclusion}. 

\section{Line lists}\label{sec:linelists}
\subsection{Sources}\label{sec:data_sources}

HITRAN~\citep{HITRAN_2016} is a database of largely experimental data which has been measured at room temperature. For this reason, although it is often very accurate - more accurate than theoretically calculated data - it is not considered complete (many of the weaker lines in particular are missing); it is only designed for the temperature in the region of 296 K although in practice HITRAN should work satisfactorily for temperatures below this. The GEISA database~\citep{GEISA} has similar properties to
HITRAN.
In order to characterise high temperature atmospheres, such as those of exoplanets and stars, theoretical calculations need to be utilised in order to compute energy levels up to high energies, along with the Einstein-A coefficients between them (giving the probability of a transition between two states). Large projects and associated databases which contain line list data appropriate up to much higher temperatures (at least 1000~K) have therefore been set up to this effect. 
These include ExoMol~\citep{jt528} \citep{jt631}, HITEMP~\citep{HITEMP,jt763}, MoLLIST~\citep{MOLLIST},  along with TheoReTS~\citep{TheoReTS}, SPECTRA~\citep{spectra}, MeCaSDA and ECaSDa~\citep{13BaWeSu.db}. It should be noted that a small number of molecules in the HITRAN database (HF, HCl, HBr, HI, H$_2$) are considered applicable up to high temperatures of around 4000~-5000~K~\citep{13LiGoHa.hitran}. 
The data for this work has been 
sourced mainly though the ExoMol database, with additions from HITEMP and MoLLIST (and HITRAN for the above mentioned molecules whose data is appropriate up to 5000~K) where data is not currently available from ExoMol, or where it is more complete and therefore recommended for use from a different source.  Atomic data is taken from the NIST database~\citep{NISTWebsite}, with original doublet data measured by \cite{81JuPiHa.Na} and \cite{06FaTiLi.K}.

\subsection{Data Format}\label{sec:data_format}

As stated above, the main data sources for this work are from ExoMol,  HITEMP/HITRAN, and MoLLIST. 
The data format of ExoMol line lists is explained in detail by \cite{jt548,jt631}, with a 2020 update in \cite{jt804}. A summary of the format is given below:
\begin{itemize}
	\item a ``.states'' file, giving the molecule's unique set of energy levels, along with a full set of quantum numbers 
	\item a ``.trans'' file, giving the transition probabilities between allowed energy states, in the form of Einstein-A coefficients
	\item (for some molecules) a ``.broad'' file, describing the broadening parameters (typically for self, air, H$_2$ or He broadening) for the transitions, as a function of rotational quantum number $J$
	\item a ``.pf'' file, giving the temperature dependent partition function 
\end{itemize}

The HITRAN and HITEMP data format, on the other hand, consist of one transition file per molecule with each line in the file representing one transition (i.e one line of the spectra). All the quantum numbers for the upper and lower energy states are contained on this same line, along with any broadening parameters. This format works for HITRAN and HITEMP databases due to their much smaller size in comparison to ExoMol line lists. The number of lines in an ExoMol line list is typically many billion for  larger polyatomic molecules, and therefore it not feasible to store such a large amount of data in HITRAN/HITEMP format. 
 
\section{Computing opacities}\label{sec:opacities}

\subsection{Cross-sections}\label{sec:xsecs}

In order to compute cross-sections from these line lists, we made use of ExoCross~\citep{ExoCross}. ExoCross is a Fortran code for generating spectra (absorption and emission) and thermodynamic properties from molecular line lists, and accepts several formats including those of ExoMol and HITRAN/HITEMP. It produces cross-sections at a specified resolution (or number of points), and broadening parameters can be included, with a variety of line broadening schemes available (see Section~\ref{sec:broad}). The MoLLIST data used in this work were converted to ExoMol format by \cite{jt790}. ExoCross is also capable of working with the super-lines method of \cite{TheoReTS}; see Section~\ref{sec:superlines}.

The partition functions for those species with data from HITRAN or HITEMP were mainly computed using TIPS (Total Internal Partition Sums), from \cite{TIPS2017}, while for data which was taken from the online ExoMol portal, at \url{www.exomol.com}, the partition function provided there was used.

It was found by \cite{RocchettoPhD} that while a sampling resolving power R=$\frac{\lambda}{\Delta\lambda}$=10,000 was sufficient to retrieve WFC3/HST spectra, a resolving power of at least 15,000 is needed to model JWST spectra. We have therefore used R=15,000 for the cross-section data presented in this work. This resolution is still insufficient for
high resolution Doppler shift studies, for this see the recent work
by \cite{jt782}.

\begin{figure}[h]
	\centering
	\includegraphics[width=0.75\textwidth]{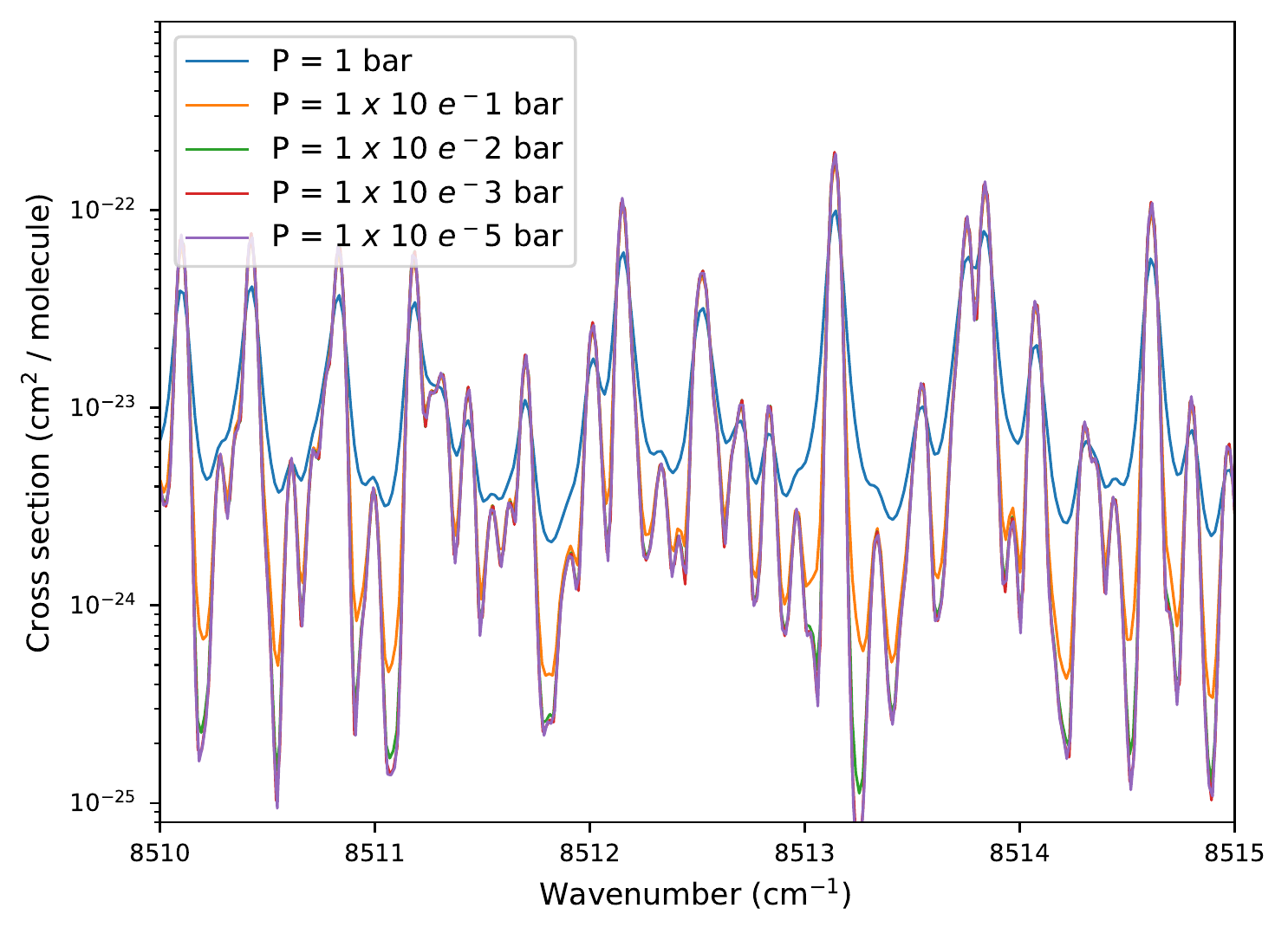}
	\caption{H$_2$O broadened by H$_2$ and He (using the parameters of Table~\ref{t:sources_2}) at a variety of pressures for T~=~1000~K. }
	\label{fig:H2O_T1000K_vary_P}
\end{figure}

\begin{figure}[h]
	\centering
	\includegraphics[width=0.75\textwidth]{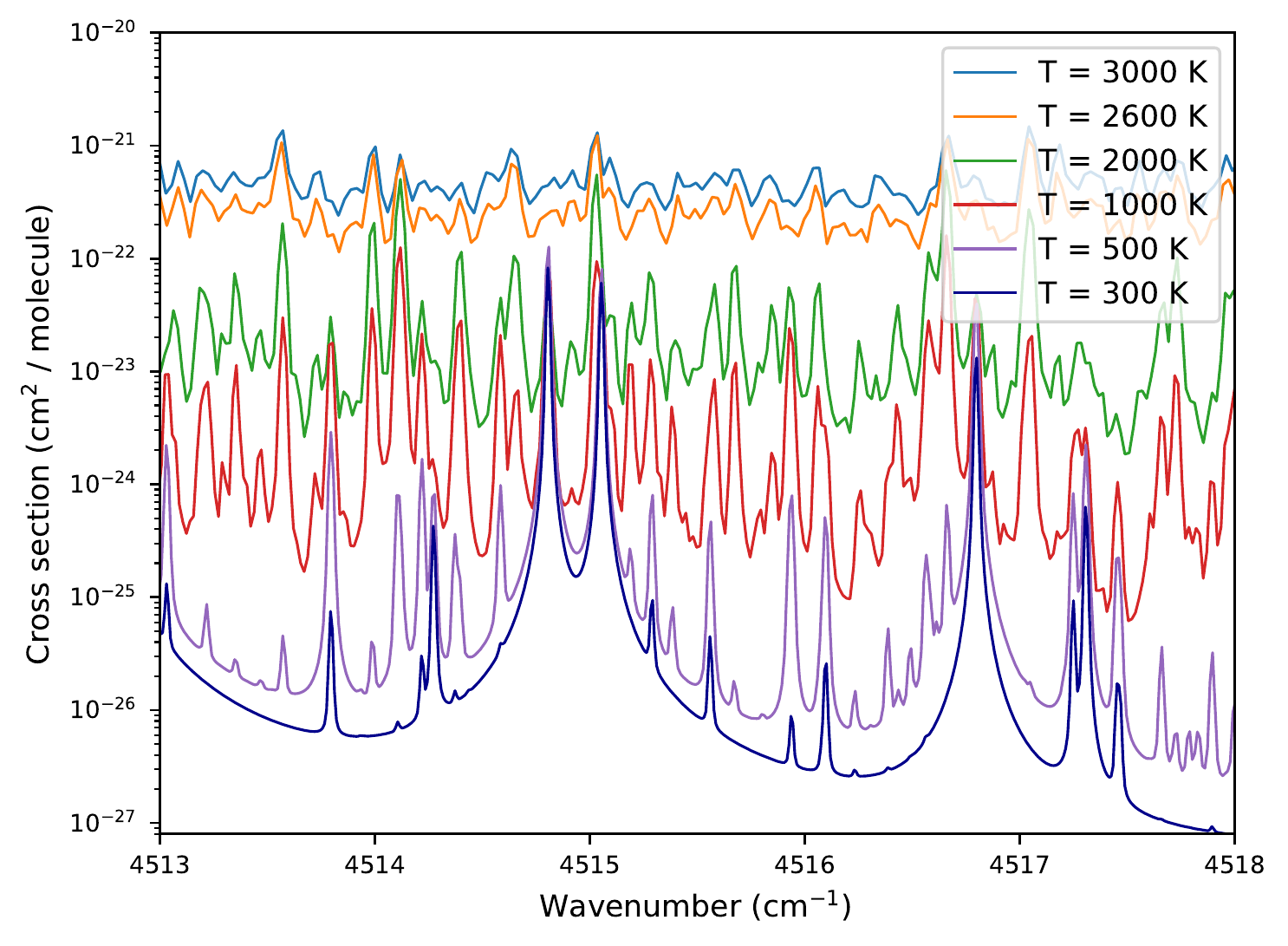}
	\caption{H$_2$O broadened by H$_2$ and He (using the parameters of Table~\ref{t:sources_2}) at a variety of temperatures for P~=~0.1~bar. }
	\label{fig:H2O_P0-1_vary_T}
\end{figure}

The cross-sections for each species are computed at the grid of 27 temperature and 22 pressures as given by Table~\ref{t:temps}, giving a total of 594 temperature-pressure combinations for each molecule. The minimum and maximum wavelength values at which cross-sections are computed is between 0.3 and 50 $\mu$m (200~-~33333~\cm), although not all molecules have this wide coverage (see Tables~\ref{t:sources_metal_oxides}~--~\ref{t:sources_ions} in Section~\ref{sec:sources}). An illustration of H$_2$O cross sections using the POKAZATEL line list~\citep{jt734} computed at a variety of pressures for T~=~1000~K and a variety of temperatures for P~=~0.1~bar are given in Figures~\ref{fig:H2O_T1000K_vary_P} and~\ref{fig:H2O_P0-1_vary_T}, respectively. The broadening parameters of Table~\ref{t:sources_2} are used, assuming a solar H$_2$:He ratio (see Section~\ref{sec:broad}). 


Different errors resulting from various aspects of computed cross-sections have been explored by \cite{16HeMaxx}, \cite{RocchettoPhD}, \cite{19GhLi.broad} and \cite{20BaChGa.exo}. 
Differences largely arise from either the choice of line list for a particular species, or how the broadening of the lines are treated, including line wing cut-offs. The differences resulting from broadening type (i.e. self- compared to H$_2$~/~He broadening) is thought to be significant for some species, such as H$_2$O, but not so much for others like CH$_4$, CO$_2$ or CO~\citep{19GhLi.broad}.

	\begin{table}[H]
	\caption{Temperatures and pressures at which the cross-sections and k-tables presented in this work are calculated at.}
	\label{t:temps} 
	\centering  
	\begin{tabular}{ccccccc}
		\hline\hline
		\rule{0pt}{3ex}Temperatures & 100 & 200 & 300 &	400 & 500 & 600  \\
		(K)	& &&&&&\\ 
		 & 700	& 800 & 900 & 1000 &	1100 & 1200  \\
		& 1300 	& 1400	& 1500 & 1600 & 1700 &	1800 \\
		&	1900 & 2000	& 2200 	& 2400 & 2600 & 2800 \\
			&	3000 & 3200 & 3400   \\
			
		\hline
			\end{tabular}
			\begin{tabular}{cccc}
		\rule{0pt}{3ex}Pressures & 1$\times$10$^{-5}$ & 2.1544$\times$10$^{-5}$ & 4.6416$\times$10$^{-5}$  \\
		(bar)	& &&\\ 
	    & 1$\times$10$^{-4}$ & 2.1544$\times$10$^{-4}$ & 4.6416$\times$10$^{-4}$ \\
	     & 1$\times$10$^{-3}$ & 2.1544$\times$10$^{-3}$ & 4.6416$\times$10$^{-3}$  \\ 
	     & 1$\times$10$^{-2}$ & 2.1544$\times$10$^{-2}$ & 4.6416$\times$10$^{-2}$ \\
	     & 1$\times$10$^{-1}$ & 2.1544$\times$10$^{-1}$ & 4.6416$\times$10$^{-1}$ \\
	      & 1 & 2.1544 & 4.6416 \\
	      & 10 & 21.544 & 46.416 \\ 
		& 100 \\
		\hline\hline
	\end{tabular}
\end{table}


Obviously a higher sampling of spectral lines leads to improvements in the accuracy of the opacity. Every line being very well sampled is equivalent to \textit{line-by-line} integration. However, sampling cross-sections at a lower resolution is far more computationally feasible for atmospheric retrievals. Retrieval results using cross-sections have been found to be generally good when the sampling resolution is around two orders of magnitude higher than the resolution of the observed spectrum \citep{RocchettoPhD}. For this work, high-resolution cross-sections were first computed for each pressure-temperature grid point (as determined by Table~\ref{t:temps}) and then sampled to a lower resolution cross-section or k-tables (see Section~\ref{sec:ktables}). The line broadening parameters used for these high-resolution cross-sections are discussed in Section~\ref{sec:broad}. For some molecules, i.e. those with many millions or billions of lines, the super-lines method was used (see Section~\ref{sec:superlines}), which has been found to vastly improve efficiency of calculations, and yields a very small error transmission provided a fine enough grid is used~\citep{jt698}. 

See Section~\ref{sec:broad} for a summary on the method used to try and ensure an adequate number of sampling points for the Voigt-broadened high-resolution cross-sections, depending on wavelength region, pressure and temperature.

\subsection{Line Broadening}\label{sec:broad}

A line list can be used to compute a simple stick spectra; a temperature-dependent list of line positions and line intensities. However, a real spectrum will always have some broadening of these spectral lines due to various processes. 
The dominant type of line broadening in an exoplanet or stellar atmosphere are Doppler (thermal) and pressure broadening. Doppler broadening, which is temperature-dependent, arises due to the thermal velocities of individual molecules, is represented by a Gaussian line profile (see, for example, \cite{ExoCross}): 

\begin{equation}\label{eq:doppler}
f_{\tilde{\nu}_{fi},\alpha_D}^D(\tilde{\nu}) = \sqrt{\frac{\ln{2}}{\pi}}\frac{1}{\alpha_D} \exp{\left( - \frac{(\tilde{\nu}-\tilde{\nu}_{fi})^2\ln{2}}{\alpha_{D}^2}\right)}, 
\end{equation}
where  $\tilde{\nu}_{fi}$ is the position of the line centre and $\alpha_D$ is the Doppler half-width at half-maximum (HWHM), as given by: 

\begin{equation}\label{eq:alpha_D}
\alpha_D = \sqrt{\frac{2k_BT\ln{2}}{m}}\frac{\tilde{\nu}_{fi}}{c}
\end{equation}
for a molecule of mass $m$, at temperature $T$, and with $k_B$ and $c$ representing the Boltzmann constant and speed of light, respectively. 

Pressure broadening, which is dependent on the perturbing species (commonly H, He, air or self) as well as the pressure, leads to a Lorentzian profile, as given by: 
\begin{equation}\label{eq:lorentzian}
f_{\tilde{\nu}_{fi},\gamma_L}^L(\tilde{\nu}) = \frac{1}{\pi}\frac{\gamma_L}{(\tilde{\nu} - \tilde{\nu}_{fi})^2 + \gamma_L^2},
\end{equation}
where the Lorentzian line width (HWHM) is given by: 
\begin{equation}\label{eq:gamma_L}
\gamma_L = \gamma_L^0 \left( \frac{T_0}{T}  \right)^{n_L} \frac{P}{P_0}.
\end{equation}
Here, $T_0$ and $P_0$ are the reference temperature and pressure, respectively. $\gamma_L^0$ and $n_L$ are the reference HWHM and temperature exponent, respectively. The latter two terms are known as pressure broadening parameters, and are dependent on the molecular species being broadened and the species causing the broadening. There is also some dependence on the rotational angular momentum quantum numbers, $J$, of the states involved in a particular transition. 

A convolution of the two profiles given in Eqs.~(\ref{eq:doppler}) and (\ref{eq:lorentzian}) gives a Voigt profile, which is commonly used to represent line broadening in exoplanet atmospheres: 
\begin{equation}\label{eq:voigt}
f_{\tilde{\nu}_{fi},\alpha_D,\gamma_L}^V(\tilde{\nu}) = \frac{\gamma \sqrt{\ln{2}}}{\pi^{\frac{3}{2}}\alpha_D} \int_{-\infty}^{\infty} \frac{e^{-y^2}dy}{(\nu - y)^2 + \gamma^2},
\end{equation}
where $\gamma = \sqrt{\ln{2}} (\gamma_L / \alpha_D)$ and $\nu = \sqrt{\ln{2}} (\tilde{\nu} - 2\tilde{\nu}_{fi}) / \alpha_D$, with terms as defined in Eqs~\ref{eq:doppler}~-~\ref{eq:gamma_L}. Technically there should be a pressure-shift of the transition wavenumber, $\tilde{\nu}_{fi}$, included in computations of line broadening. There are, however, currently large experimental uncertainties associated with the relatively small number of pressure-shift values which are currently available. There is work ongoing to improve upon these parameters \citep[e.g.][]{jt763,18GaVi.broad}, and we hope to include more accurate parameters such as these in future opacity calculations.

The efficient and accurate numerical computation of such a profile has been the subject of a number of publications \citep[see, for example,][]{16HeMaxx,17Min.methods,17Schreier,15GrHexx,ExoCross,14AmBaTr.exo}. The coefficients $\gamma_0$ and $n_L$ in Eq.~(\ref{eq:gamma_L}) are dependent both on the species which is being broadened, and on the species which are causing the broadening (for example, H$_2$ and He are assumed to be the main broadeners in typical ``Hot Jupiter'' atmospheres~\citep{16HeMaxx}), along with the rotational angular momentum quantum numbers, $J$, of the states involved in a particular transition.  

As mentioned in Section \ref{sec:xsecs}, the number of sampling points required to give well-sampled Voigt-broadened high-resolution cross-sections can be estimated as a function of wavelength $\nu$, temperature $T$, and pressure $P$, with an average number found for a given wavelength region. The Voigt width of a particular line can be approximated by the following expression~\citep{77OlLo.broad,RocchettoPhD}:
\begin{equation}\label{eq:voigt_width}
\gamma_V \approx 0.5346 \gamma_L + \sqrt{0.2166 \gamma_L^2 + \gamma_G^2},
\end{equation}
where $\gamma_L$ and $\gamma_G$ are given by Eqs.~\ref{eq:alpha_D} and \ref{eq:gamma_L}, respectively. In this work, we aim for an average of 4 sampling points per line for the initial set of high-resolution cross-sections. This is estimated using Eq.~(\ref{eq:voigt_width}), as a function of pressure, temperature and wavelength region. We use a line-wing cut-off of 500~$\gamma_V$, which is also calculated using Eq.~\ref{eq:voigt_width} and is thus dependent on pressure, temperature and wavelength.

Where available, broadening parameters ($\gamma_0$ and $n_L$ in Eq.\ref{eq:gamma_L}) are provided as part of the ExoMol database (see \cite{jt662, jt684}). However, in general, these broadening parameters are not well known (or given in the literature at all) for a large number of species, particularly for He and H$_2$ as broadeners. In this work we aim to use approximate pressure broadening parameters with He and H$_2$ as the broadening species (as they are thought to be the main constituents of ``hot Jupiter'' atmospheres), at solar H$_2$:He ratio, based upon those parameters that do exist in the literature, and using molecular properties, as given in Table~\ref{t:sources_2} and Tables~\ref{t:ch4}~-~\ref{t:hcn} to estimate which parameters to use for those which do not. Table~\ref{t:sources_2} lists those species for which broadening parameters exist in the ExoMol database (primarily based upon the work of \cite{jt684}). These parameters are usually given as a function of rotational angular momentum quantum number, $J$, for each species. Here, we have taken an average value for all values of $J$ for a given species, for broadening of both H$_2$ and He. When computing the cross sections, we then weight these parameters by assumed broadener abundances, based on the solar H$_2$:He ratio. 
When deciding which species is most similar from a broadening point of view to those in Table~\ref{t:sources_2}, we consider the following factors.  
First we considered the dipole moment (DM), and quadrupole moment (QM) where DM=0~\citep{10BuLaSt.broad}. We then also look at the general structure (e.g. linear, non-linear, diatomic), and consider molecular properties such as the centre of symmetry and interatomic distances.  Of the nonpolar (DM=0) species in Table~\ref{t:sources_2}, CH$_4$ has QM=0, H$_2$ has a small QM, and C$_2$H$_2$ has a relatively large QM. The other nonpolar molecules are grouped accordingly, based on whether their QM is zero, low or high. The reference for the value of the dipole moment in these tables is given where applicable. For those with no obvious value in the literature, the dipole moment was computed using MOLPRO~\citep{MOLPRO}, with a cc-pVTZ basis set at CASSCF level of theory~\citep{Olsen2011}. Where multiple values of dipole moment are given in the literature (depending on the level of theory used to calculate it), we used an average value. It is stressed that these values are only given as a guide towards determining which species in Table~\ref{t:sources_2} is most similar to those in Tables~\ref{t:ch4}~-~\ref{t:hcn}, and should not be taken as exact. For those diatomic species in Table~\ref{t:sources_2} where line-by-line broadening parameters are provided (for example, by \cite{15LiGoRo.CO,16WiGoKo.pb, HITRAN_2016}), we use these values instead of the average parameters when computing their opacities.

\begin{figure}[h]
	\centering
	\includegraphics[width=0.75\textwidth]{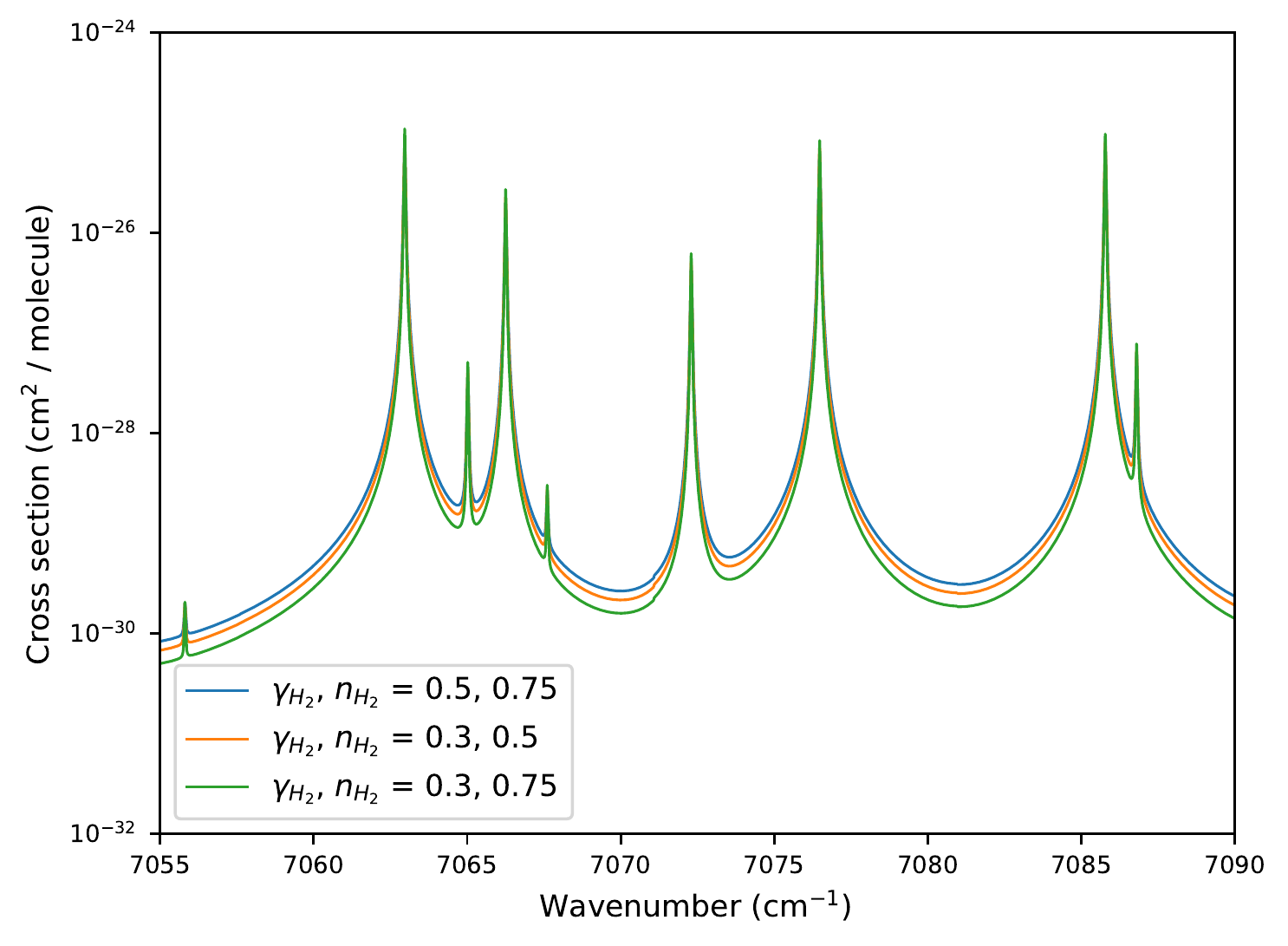}
	\caption{HCl broadened by H$_2$ using different broadening parameters. 
		The cross-sections are computed at T~=~1000~K and P~=~0.1~bar, which are typical values for a layer of atmosphere of a ``Hot Jupiter'' exoplanet being observed in this wavelength region. }
	\label{fig:HCl_broadening}
\end{figure}

Figure~\ref{fig:HCl_broadening} illustrates the effect of using different broadening parameters for HCl when computing the cross sections for a pressure and temperature typical of a ``Hot Jupiter'' exoplanet being observed in the given wavelength region. The broadening is considered to be from H$_2$ only in this example. 

Studies such as those by \cite{16HeMaxx}, \cite{RocchettoPhD}, \cite{17BaMoVe.exo}, \cite{19GhLi.broad} and \cite{20BaChGa.exo} have highlighted the differences in forward models and retrievals caused by the choice of molecular broadening parameters.
 It is important to note that although there is a known difference between different parameters, there is still much work required in order to determine the ``true'' parameters, or even a good approximation, for many species. The requirement for enhanced line broadening parameters was recently identified by \cite{jt773} in their study on the need for laboratory data requirements for studies of exoplanetary atmospheres.
 This is the focus of various works 
  such as \cite{20StThCy.broad,18HaTrAr.broad}, who are improving upon the current knowledge of molecular broadening. Some detailed accounts of line-broadening theory can be found in, for example, \cite{10BuLaSt.broad,10BuLaStb.broad} and \cite{16WcGoTr.H2}. The latter demonstrates that H$_2$ has ``exceptionally pronounced non-Voigt line-shape effects''. We do not take these more precise effects into account in this work, but they could be considered in future work. 
 Other intricacies have been neglected here.
 We note, for example, that broadening parameters are not only dependent on $J$, but also on the lower state symmetry (see, for example, \cite{19GhHeBe.ch4}). 
The main focus of this work, however, is towards useable opacities for low-resolution atmospheric retrieval studies. 
There are many other limiting factors when it comes to detecting molecules and making accurate determinations of their abundances. The accurate treatment of pressure effects is beyond the scope of the current work, but is planned to be significantly enhanced in future releases of the ExoMol and therefore ExoMolOP database.


	
	\begin{table*}[h]
		\caption{Molecular properties for the species with computed opacities presented in this work for which H$_2$ and He broadening parameters are available in the literature, the majority of which were collated by \cite{jt684}. $\gamma_x$ and $n_x$ (where x=H or He) are the broadening parameters for hydrogen and helium ($\gamma_0$ and $n_L$ in Eq.\ref{eq:gamma_L}). DM is the dipole moment, MM is the molar mass, and AN is the atomic number. The citations for the H$_2$~/~He broadening parameters are given below the table for each molecule. An average value for all values of $J$ is taken for each set of broadening parameters. The reference temperature of $T_0$~=~296~K is used in all cases.}
		\label{t:sources_2} 
		\centering  
		\begin{tabular}{llllllllll}
			\hline\hline
			\hline
\rule{0pt}{3ex}Species	&	$\gamma_{H_2}$	&	$n_{H_2}$	&	$\gamma_{He}$	&	$n_{He}$	&	DM	&	MM	&	AN	&	Structure	& Dipole Ref		\\
	&		&		&		&		&	 (d)	&	(g/mol)	&		&				\\
			\hline\hline
\rule{0pt}{3ex}H$_2$	&	0.01	&	0.13	&	0.01	&	0.13	&	0	&	2.0	&	2	&	Nonpolar	& \cite{CCCBDBWebsite} 		\\
CH$_4$	&	0.06	&	0.60	&	0.03	&	0.30	&	0	&	16.0	&	10	&	Nonpolar			\\
C$_2$H$_2$	&	0.09	&	0.59	&	0.04	&	0.44	&	0	&	26.0	&	14	&	Linear			\\
CO	&	0.07	&	0.65	&	0.05	&	0.60	&	0.1	&	28.0	&	14	&	Diatomic	& 	\cite{CCCBDBWebsite}		\\
PH$_3$	&	0.10	&	0.75	&	0.05	&	0.30	&	0.6	&	34.0	&	18	&	Non-linear			\\
OCS	&	0.05	&	0.75	&	0.04	&	0.75	&	0.7	&	60.1	&	30	&	Linear		& \cite{69DiRu.OCS}	\\
HCl	&	0.03	&	0.75	&	0.01	&	0.75	&	1.1	&	36.5	&	18	&	Diatomic	& \cite{CCCBDBWebsite}			\\
NH$_3$	&	0.08	&	0.50	&	0.03	&	0.50	&	1.5	&	17.0	&	10	&	Non-linear			\\
SO$_2$	&	0.14	&	0.75	&	0.07	&	0.64	&	1.6	&	64.1	&	32	&	Non-linear			\\
H$_2$O	&	0.06	&	0.20	&	0.01	&	0.13	&	1.9	&	18.0	&	10	&	Non-linear			\\
HF	&	0.04	&	0.75	&	0.01	&	0.50	&	1.9	&	20.0	&	10	&	Diatomic	& \cite{CCCBDBWebsite}			\\
H$_2$CO	&	0.14	&	0.50 &	0.06	&	0.50	&	2.3	&	30.0	&	16	&	Non-linear			\\
HCN	&	0.12	&	0.50	&	0.05	&	0.50	&	3.0	&	27.0	&	14	&	Linear	& \cite{TomasevichPhD}		\\
			\hline\hline
		\end{tabular}
	\mbox{}\\
	
	{\flushleft
		H$_2$: \cite{16WcGoTr.H2}.\\
		CH$_4$: \cite{90VaChxx.CH4,92Pine.CH4,98FoJeSt.CH4,72VaTexx.CH4,89VaChxx.CH4,01GrFiTo.CH4,04GaGrGr.CH4,17MaBuWe,19GhHeBe.ch4} \\
		C$_2$H$_2$, OCS, NH$_3$, SO$_2$, HF: \cite{16WiGoKo.pb} \\
		HCl: \cite{16WiGoKo.pb,18LiAsDo.broad}\\ 
		CO: \cite{jt544,15LiGoRo.CO,18MuSuAl.co,05MaDeMa.co,16PrEsRo.co,98SiDuBe.co} \\
		PH$_3$: \cite{03KlTaCo.PH3,93LeLaTa.PH3,88SevaRo.PH3,04SaBoWa.PH3,06SaBoWa.PH3}  \\
		H$_2$O: \cite{jt483,12VoLaLu.H2O,jt669,08SoStxx.H2O,09SoStxx.H2O,13PeSoSo.H2O,12PeSoSt.H2O,16PeSoSo.H2O,19GaViRe.h2o} \\
        H$_2$CO: \cite{75Nerf.H2CO} \\
        HCN: \cite{85MeMaVr.HCN,73CoWixx.HCN,80ChAnSt.HCN} \\
	}
	\end{table*}



\subsubsection{H$_2$S: A broadening case study} 

We can compare the molecular properties of H$_2$S to those of the species listed in Table~\ref{t:sources_2} which have some broadening parameters available in the literature. The dipole moment of H$_2$S is closest to HCl. However, the intermolecular distances and potential energy surface of H$_2$S are more similar to OCS than HCl~\citep{CCCBDBWebsite}, so it is not clear which broadening parameters should be used. Here, we compute opacities using the two different sets of broadening parameters ($\gamma_{H_2}$, $n_{H_2}$, $\gamma_{He}$, $n_{He}$ = 0.03, 0.75, 0.01, 0.75 and 0.05, 0.75, 0.04, 0.75, for HCl and OCS, respectively) in order to illustrate the effects of using different broadening parameters. 
For HCl, the average values of the broadening parameters are used from \cite{16WiGoKo.pb}, with the exception of $n_{H_2}$. This parameter was originally sourced by \cite{16WiGoKo.pb} from the work of \cite{80HoLaHa.broad}, which presents negative temperature dependence exponents. As noted by \cite{16WiGoKo.pb}, negative exponents are not impossible but they are unexpected, particularly for a simple diatomic like HCl. Negative exponents for HCl-$N_2$ were also previously published by \cite{80HoLaHa.broad}, which were subsequently found to be positive by \cite{87PiLo.broad}. This gives further cause to be cautious about using the negative values. We therefore instead use the suggested default value of $n_{H_2}$~=~0.75 for HCl. There are some values for $\gamma_{He}$ \citep{94WaKuSu.broad} and $\gamma_{H_2}$ \citep{02KiSuKr.broad,06StPr.broad} for the $\nu_2$ vibrational band of H$_2$S. These range from 0.04~-~0.07 for $\gamma_{H_2}$ and 0.04~-~0.1 for $\gamma_{He}$, which give further agreement with the values of OCS over those of HCl. 

\begin{figure}[h]
	\includegraphics[width=0.5\textwidth]{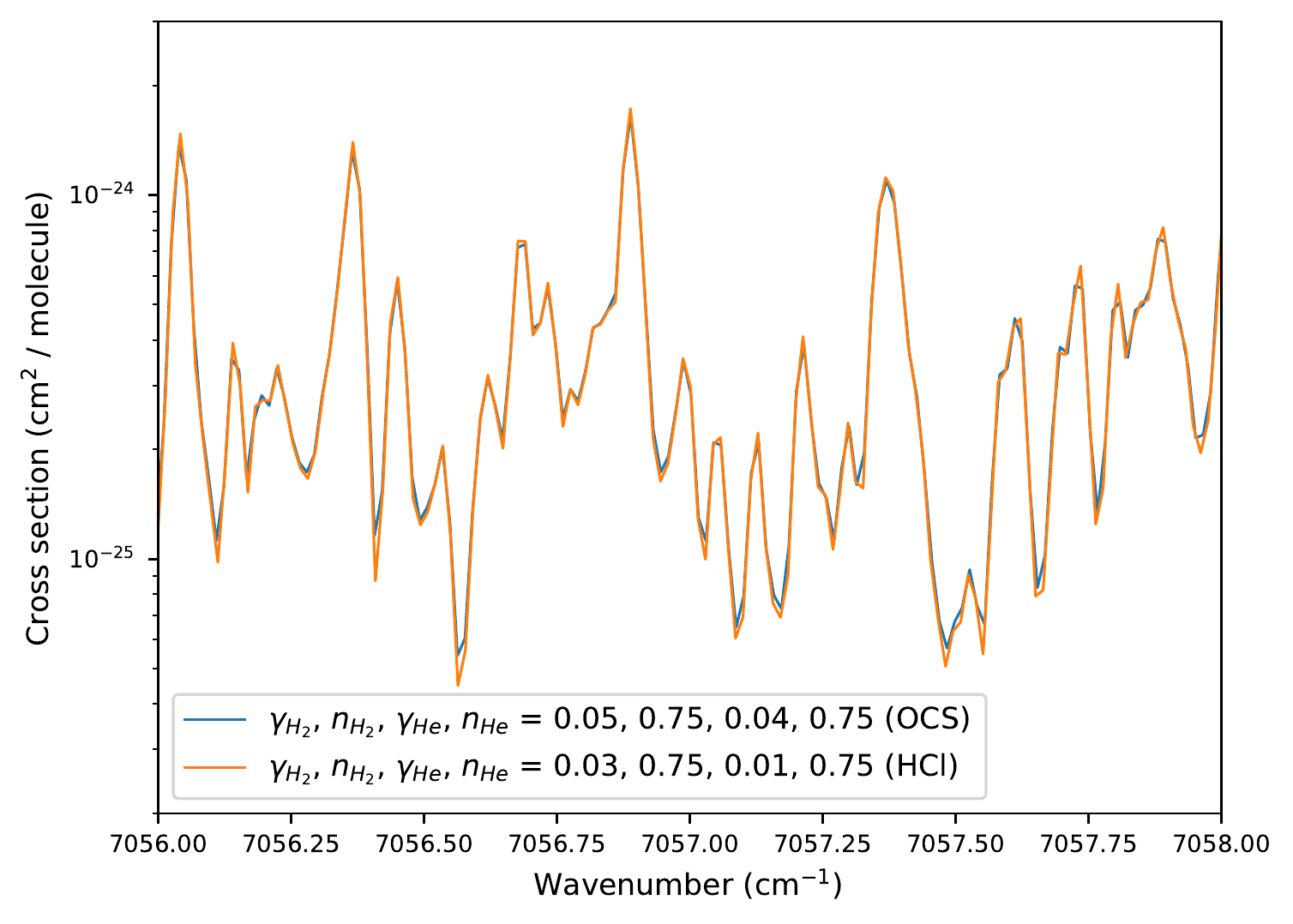}
	    \includegraphics[width=0.5\textwidth]{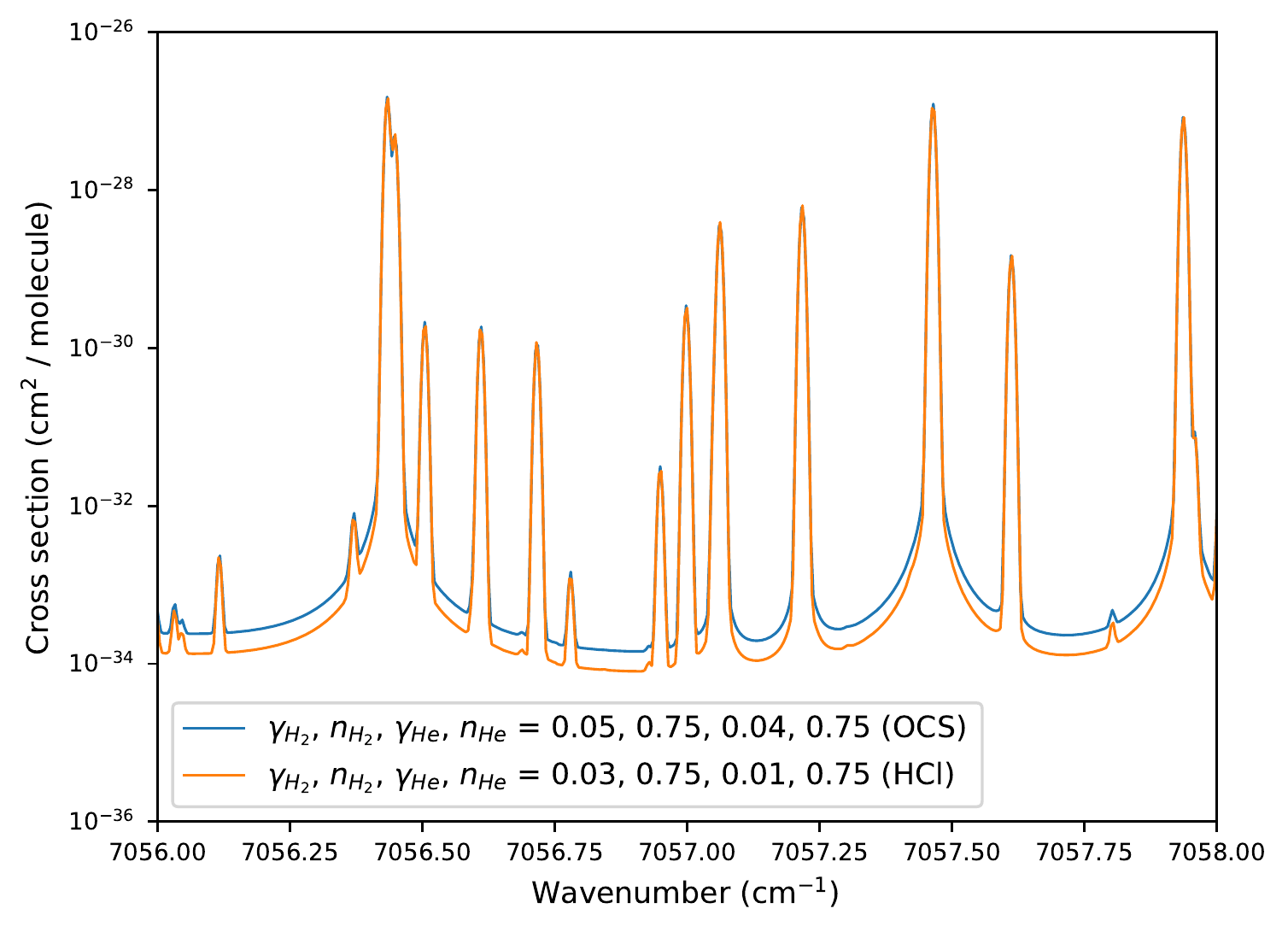}
	\caption{H$_2$S broadened by H$_2$ and He using different broadening parameters, based on values for HCl and OCS. 
		The cross sections in the left panel are computed at T~=~1000~K and (P~=~0.1~bar), which are typical values for a layer of atmosphere of a ``Hot Jupiter'' exoplanet being observed in this wavelength region. The cross sections in the right panel are computed at T~=~100~K and P~=~1~$\times$~10$^-5$~bar.}
	\label{fig:H2S_broadening}
\end{figure}

\begin{figure}[h]
	\includegraphics[width=0.5\textwidth]{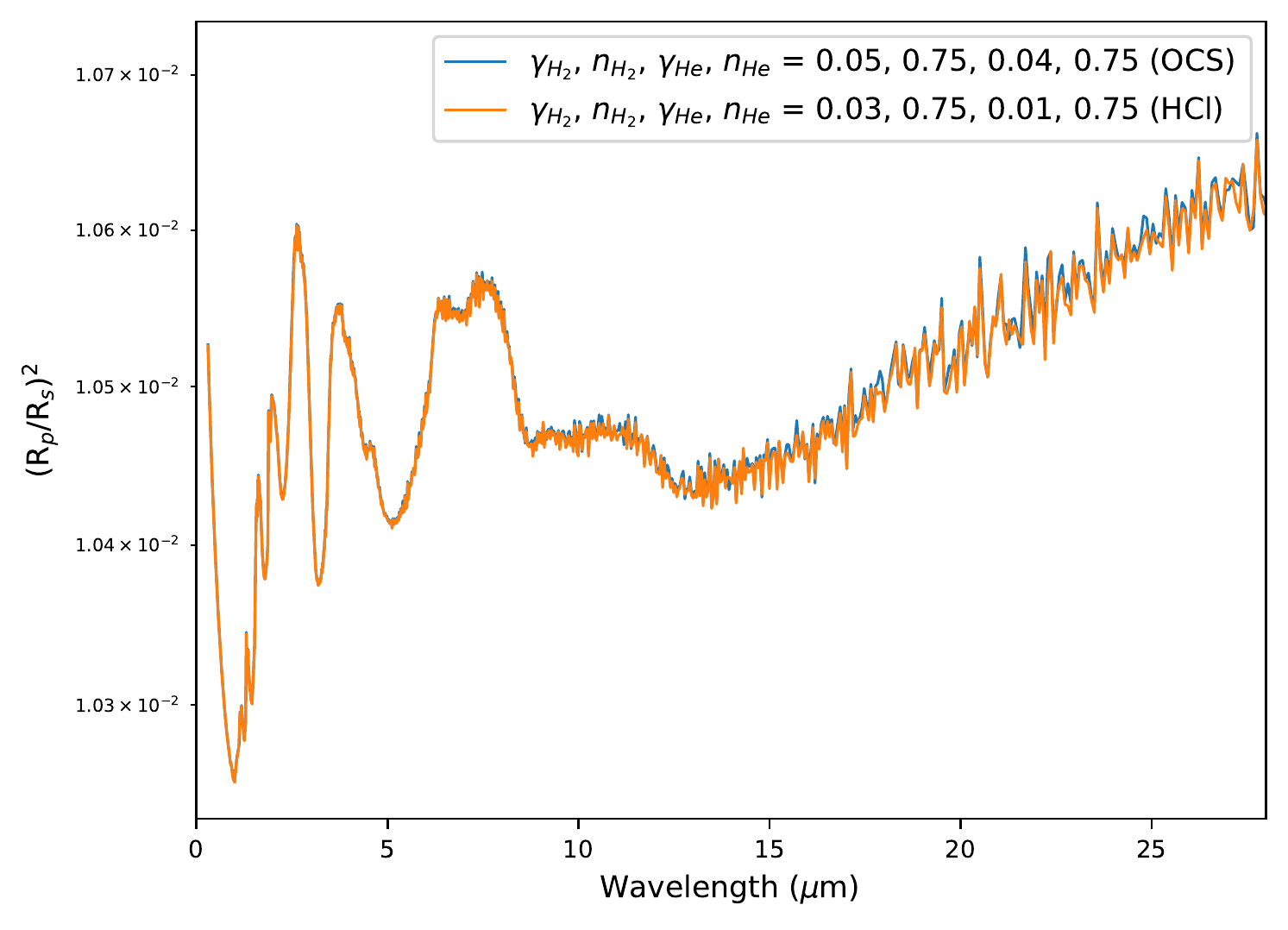}
	\includegraphics[width=0.5\textwidth]{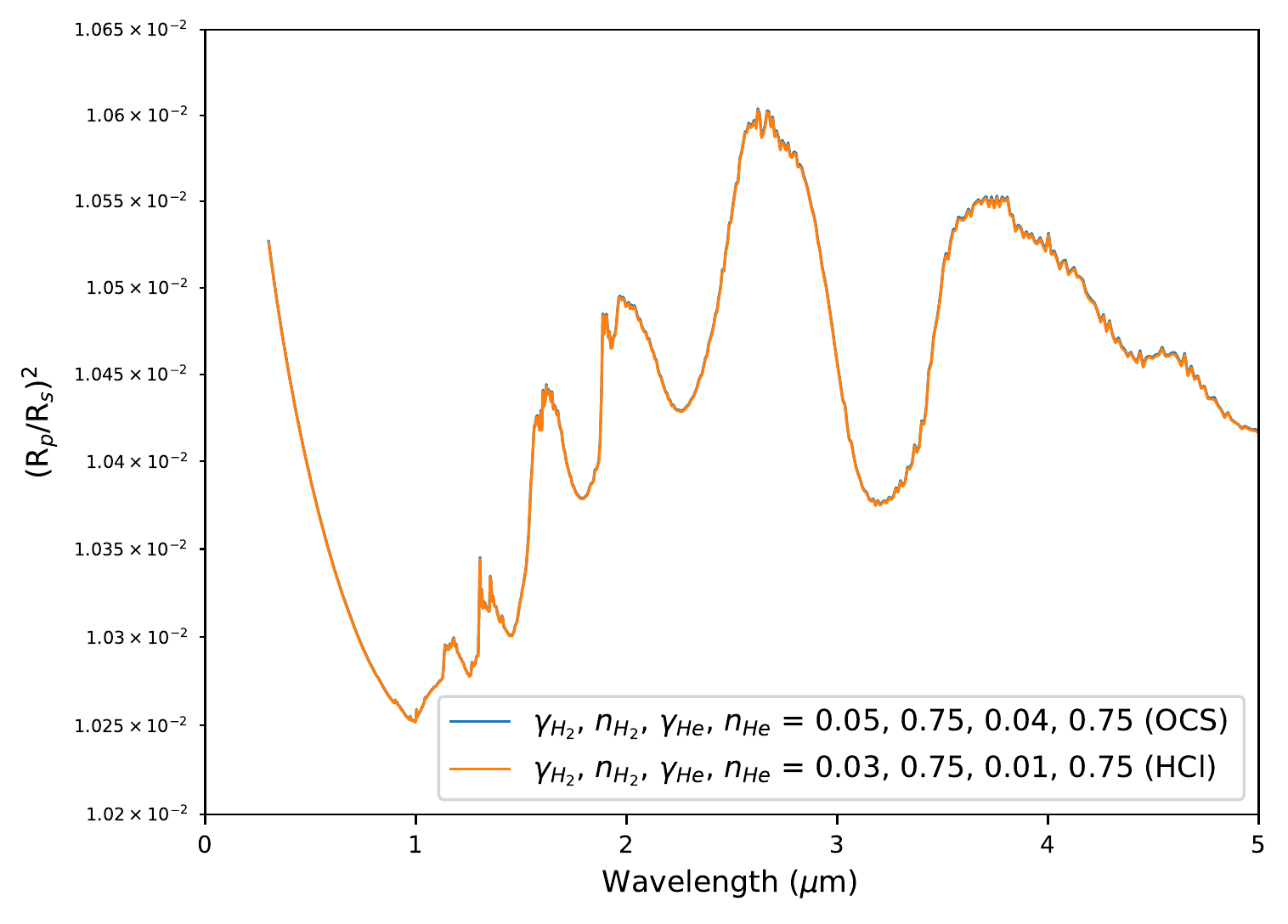}
	\caption{A hypothetical model atmosphere of an exoplanet, composed of H$2$S only. H$_2$S is broadened by H$_2$ and He using different broadening parameters in the two cases shown in each panel, based on values for OCS and HCl. 
		The model atmosphere is computed at T~=~1000~K across pressures ranging from 1~$\times$~10$^{-5}$ to 1~$\times$~10$^{6}$ bar. The two panels illustrate different regions of the spectrum. }
	\label{fig:H2S_broadening_atm}
\end{figure}

Figure~\ref{fig:H2S_broadening} illustrates the difference in computed cross sections of H$_2$S between using the broadening parameters of Table~\ref{t:sources_2} for OCS and HCl. A couple of different pressure-temperature combinations are used for comparison. It can be seen that the difference is more pronounced for lower temperatures and pressures, as was found in previous studies~\citep{16HeMaxx,RocchettoPhD}. Figure~\ref{fig:H2S_broadening_atm} illustrates forward models of a hypothetical atmosphere (computed using TauREx~\citep{15WaTiRo.taurex,15WaRoTi.taurex}), comparing opacities computed using these two different sets of broadening parameters. The model atmosphere is computed at 1000~K and contains H$_2$S only. It can be seen that the differences are small, but more pronounced at higher wavelengths. For the pressures and temperatures typical of a ``Hot Jupiter'' exoplanet observed in the near infrared, we do not expect the difference in broadening parameters to have a significant effect. Other uncertainties caused by, for example, incomplete line lists or line wing cutoffs, have been demonstrated by studies such as~\cite{RocchettoPhD} to have a larger effect on the atmospheric spectrum. The differences in opacities caused by the use of different broadening parameters, however, should always be taken into consideration when interpreting and discussing results. 


\subsubsection{Comparison of broadening values used here to other works} 

There can be found in the literature some broadening parameters for a small number of other species, which can then be compared to those we are using in this work. A couple are given here. 

The value of $\gamma_{H_2}$ for C$_2$H$_4$ from \cite{92BrVa.broad} is $\sim$0.12~cm$^{-1}$atm$^{-1}$. We assume a value of $\gamma_{H_2}$~=~0.09~cm$^{-1}$atm$^{-1}$, based on the broadening parameters of C$_2$H$_2$ from Table~\ref{t:sources_2}. 

\begin{figure}[h]
	\includegraphics[width=0.5\textwidth]{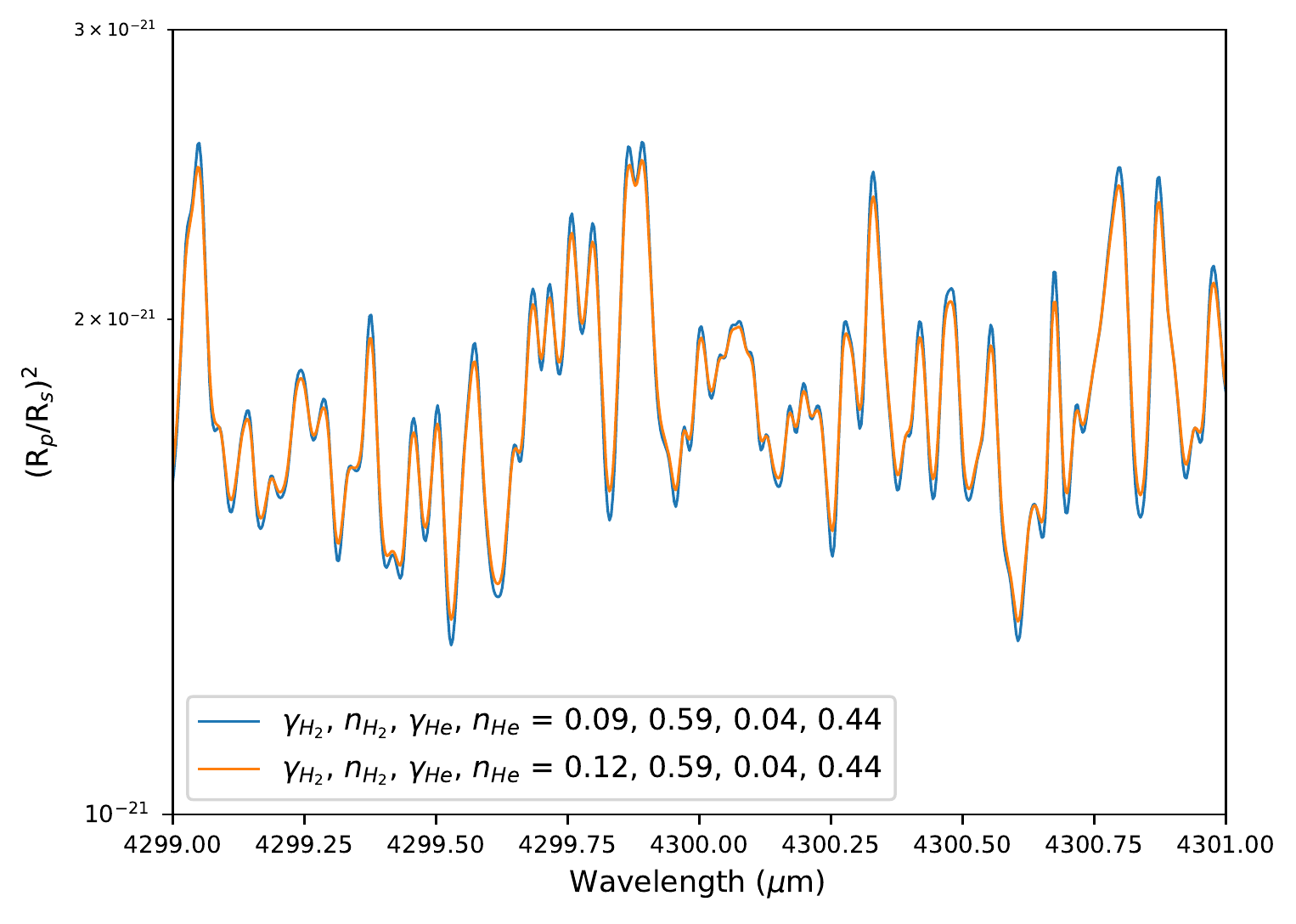}
	\includegraphics[width=0.5\textwidth]{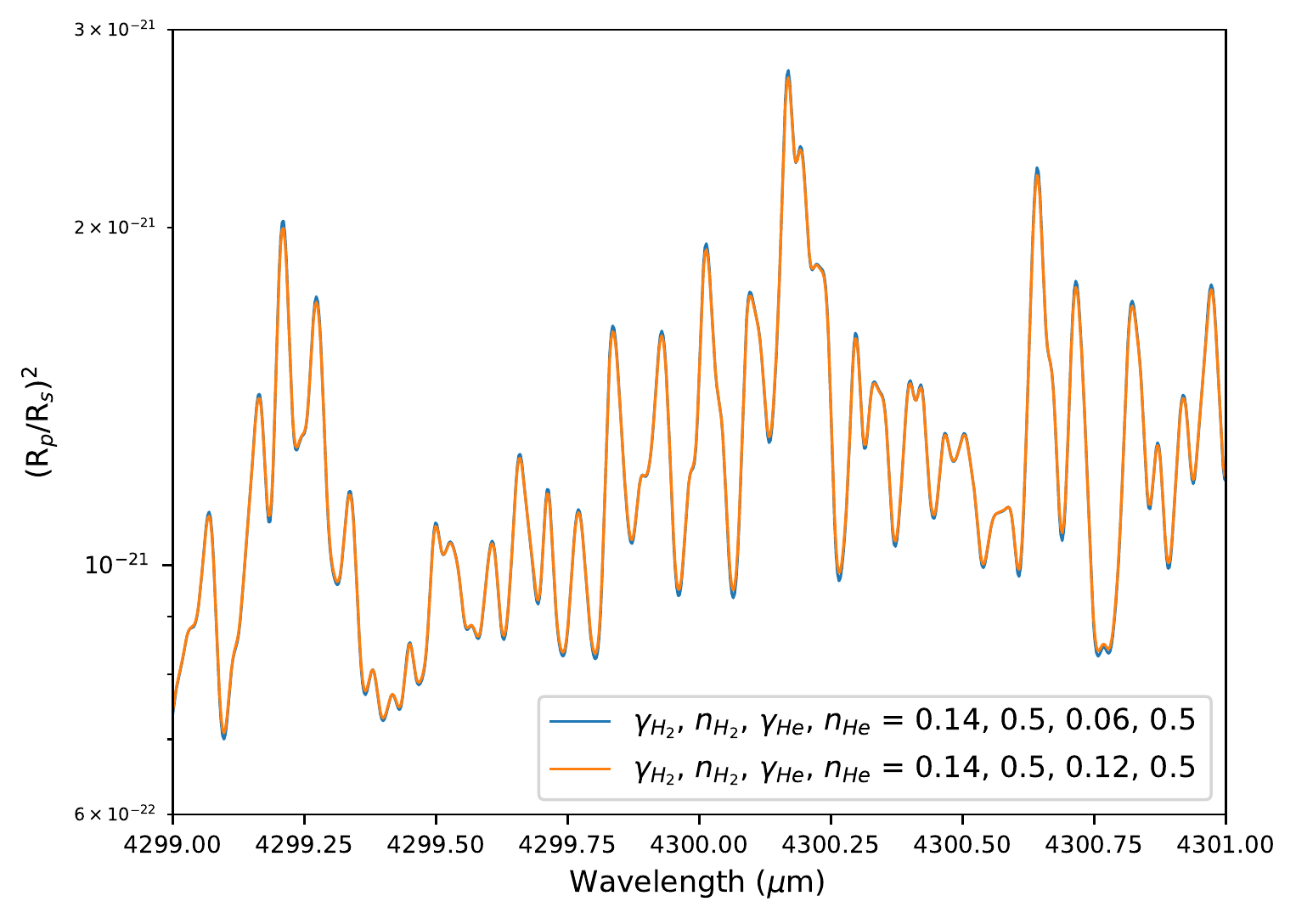}
	\caption{C$_2$H$_4$ (left panel) and CH$_3$F (right panel)broadened by H$_2$ and He using different broadening parameters. 
		The cross-sections are computed at T~=~1000~K and P~=~0.1~bar, which are typical values for a layer of atmosphere of a ``Hot Jupiter'' exoplanet being observed in this wavelength region. }
	\label{fig:C2H4_CH3F_broad}
\end{figure}

The values of $\gamma_{H_2}$ and $\gamma_{He}$ for CH$_3$F from \cite{06LeGh.broad} and \cite{97GrBoBl.broad} are $\sim$0.14~cm$^{-1}$atm$^{-1}$ and $\sim$0.12~cm$^{-1}$atm$^{-1}$, respectively. We assume values of $\gamma_{H_2}$~=~0.14~cm$^{-1}$atm$^{-1}$ and $\gamma_{He}$~=~0.06~cm$^{-1}$atm$^{-1}$, based on the broadening parameters of H$_2$CO from Table~\ref{t:sources_2}. Figure~\ref{fig:C2H4_CH3F_broad} illustrates the effects on cross-sections computed using these different broadening parameters for C$_2$H$_4$ (left panel) and CH$_3$F (right panel). We note that the temperature exponents are not given in the literature for these species, so we do not have a full set of broadening parameters to add to Table~\ref{t:sources_2}.

\cite{16HeMaxx} use the following metric to quantify the effect of using different broadening parameters, which we adopt here. 
\begin{equation}
\delta = median (\frac{\sigma - \sigma_0}{\sigma_0}) \times 100.
\end{equation}
$\delta$ is therefore the median percent change in cross section (computed at a given pressure $P$ and temperature $T$), with $\sigma$ and $\sigma_0$ are the cross sections computed using two different sets of broadening parameters. Here we compute $\delta$ for cross sections of CO using the $J$-averaged broadening parameters of Table~\ref{t:sources_2} compared to cross sections of CO using the $J$-dependent broadening parameters of \cite{15LiGoRo.CO}. Here we are comparing the high resolution cross sections, before sampling to k-tables or cross sections of lower resolution. We find that $\delta$ is 0.16$\%$ for P~=~0.1~bar and T~=~1000~K, and that $\delta$~$<$~0.3$\%$ for all pressure/temperature combinations considered in this work. $\delta$ is highest for the lowest values of pressure and temperature. \cite{16HeMaxx} find that the effect of using different broadening parameters is more pronounced at higher resolution. We therefore do not expect the approximations of broadening parameters used in this work to be a significant issue for low-resolution studies, particularly for high temperatures ($>$~1000~K) and pressures ($>$~1~$\times$$10^{-3}$~bar). We are aware, however, that theoretical and observational advances in the near future will mean that it will be beneficial to update ExoMolOP with more accurate parameters. 


\subsubsection{Atomic Na and K broadening}\label{sec:atomic_broad}


While the broadening parameters for molecular species are currently largely uncertain, the same is not true for the strong doublet lines in alkali metals sodium and potassium. These strong absorption features can be found in many high-temperature atmospheres; ``hot Jupiters''~\citep{17LeCuHa.exo,14SiWaSh.exo}, stellar atmospheres~\citep{12TaKaHa.exo}, and Brown Dwarfs, where they have been observed to be non-Lorentzian \citep{00BuMaSh.exo,02BuBuKi.exo}. Studies such as these have motivated the use of more detailed quantum chemistry calculations, which have been employed in order to accurately treat the broadening of these lines by H$_2$ and He for a variety of pressures and temperatures~\citep{09PeWi.broad,07AlKiAl.alkali,03BuVo.broad,17Peach.broad,19AlSpLe.broad}.

In this work, the pressure and temperature broadened profiles for the resonance doublets of Na and K are computed using \cite{16AlSpKi.broad} and \cite{19AlSpLe.broad}. Future updates to these opacities intend to also consider He-broadening, as outlined in~\cite{20PeYuCh.broad}.  The data for all other lines is taken from NIST~\cite{NISTWebsite}, with Voigt broadening based on parameters inferred from \cite{07AlKiAl.alkali}, and line-wing cutoffs computed at 4500\cm, as recommended by \cite{17BaMoVe.exo}. We do note that computing line wings out to this distance may not be wise when the profiles are not accurately known. However, preliminary tests indicate that the chosen line wing cut-off for these non-resonance K lines makes no noticeable difference to model emission spectra of Brown Dwarf atmospheres (where the differences are more pronounced than in exoplanet atmospheres), as the non-resonant K lines were drowned out by other opacities. The same is assumed to be true for Na.


\subsubsection{Super-lines}\label{sec:superlines}

The super-lines method \citep{TheoReTS} is a new functionality of ExoCross~\citep{ExoCross} which allows for a huge increase in computational speed and therefore efficiency. The general principle is to compute line intensities at a given temperature $T$, and then to sum together the intensities of all lines within a specified spectral bin to create a so-called super-line. A Voigt broadening profile is then applied to each of these super-lines. As long as each of these super-lines is adequately sampled then overall opacity will be conserved. We use this method when computing opacities of larger molecules, which contain many millions to billions of lines; see Tables~\ref{t:sources_triatomics} and \ref{t:sources_larger_molecules}. In order to ensure error transmission is kept to a minimum, we use a grid of R~=~1,000,000 for the super-lines~\citep{jt698,jt804}.



\subsection{K-tables}\label{sec:ktables}

Once relatively high-resolution cross-sections have been computed, it is reasonably simple to produce k-tables, using what is known as the correlated-k distribution method; this method is well described in the literature (see, for example, \cite{91LaOi.ktab,10PierreH.exo}), and is extensively used for radiative transfer calculations in the context of planetary and substellar atmospheres \citep[see, for example,][]{NEMESIS,09ShFoLi.exo,08FrMaLo.exo,14FrLuFo.exo,19LeTaGr.exo,15MoBoDu.petitCODE,14AmBaTr.exo,07ShBu.exo,17MaGrMe.exo,16DrTrBa.exo,20PhTrBa.exo}. 
k-tables are generally considered faster (and more accurate, for the same R~=~$\frac{\lambda}{\Delta \lambda}$) than cross-sections, but they also come with their own assumptions and therefore limitations \citep[see, for example,][]{RocchettoPhD}. These, however, can be thought to be negligible when compared to other unknowns. The general principle is to order spectral lines within a given spectral bin, producing a smooth cumulative distribution function to represent opacity, which can be more efficiently sampled. The number of points used for sampling within a given spectral bin is determined by a set of Gaussian quadrature points, which are assigned corresponding weights. These are often chosen so as to sample the extremes of the bin more finely so as not to miss the weakest and strongest lines (i.e. the distance between sampling points within a bin is not constant). One of the methods used in this work, for example, is based on the use of Gauss Legendre polynomials (see Section~\ref{sec:ret_codes} for details for the opacities produced for individual retrieval codes). k-tables are produced using a method of opacity sampling which enables low resolution computations while still taking strong opacity fluctuations at high resolution into account; see, for example, \cite{17Min.methods}. The assumption made for the k-distribution method, that the k-coefficients at each Guassian quadrature point are correlated, breaks down for inhomogeneous atmospheres~\citep{19LeTaGr.exo}



\section{Retrieval Codes}\label{sec:ret_codes}

There are several exoplanet retrieval codes in use by the exoplanetary characterisation community, with the aim to solve the radiative transfer equation, which looks at the propagation of radiation through a medium. We have tailored the data computed as part of this work to be directly available in the necessary format for four such retrieval codes, with no conversion necessary. These codes are ARCiS\footnote{\url{https://www.exoclouds.com}}~\citep{20MiOrCh.arcis}, TauREx\footnote{\url{https://taurex3-public.readthedocs.io}}~\citep{15WaTiRo.taurex,15WaRoTi.taurex,19AlChWa.taurex}, NEMESIS~\citep{NEMESIS}, and petitRADTRANS~\citep{19MoWaBo.petitRADTRANS}.

 A summary of each code is given in this section, along with the specific requirements of the data format of the input opacity files (cross-sections for TauREx, and k-tables for the others) for each.


\subsection{TauREx}\label{sec:taurex}

TauREx is a modular Bayesian inverse retrieval suite optimised for speed (on CPU and GPU platforms) and ease of use. It was originally designed for retrievals of exoplanet transmission, emission and phase-curve measurements \citep{19AlChWa.taurex,20Chetal}, but has recently been extended to solar system measurements \citep[e.g. ExoMars Trace Gas Orbiter (TGO),][]{20Cannal}. TauREx is publicly available\footnote{\url{https://github.com/
ucl-exoplanets/TauREx3_public}} under a BSD license.

TauREx3 has recently been released by \cite{19AlChWa.taurex}, with vast speed improvements compared to previous versions of the code~\citep{15WaTiRo.taurex,15WaRoTi.taurex}. The currently available version of TauREx3 only has support for cross-section opacities using HDF5, pickle and Exo-Transmit~\citep{17KeLuOw.exo} formats. The next release (version 3.1) will include k-table support for both petitRADTRANS and NEMESIS formats. 
The cross-section data can either be streamed directly or loaded into memory. The data contained in the HDF5 file is summarised in Table~\ref{t:taurex_hdf5}. 

\begin{table}[h]
	\caption{Overview of the data fields contained within the HDF5 cross-sections for use in the TauREx retrieval code.}
	\label{t:taurex_hdf5} 
	\centering  
	\begin{tabular}{ll}
		\hline\hline
		\hline
		\rule{0pt}{3ex}Field name & Description	\\
		\hline\hline
		\rule{0pt}{3ex}$mol\_name$	& Molecule name \\
		$key\_iso\_ll$	& ID for isotopologue and line list	\\
		$t$	& List of temperatures	\\
		$p$	& List of pressures	\\
		$t.units$	& Units of temperature (K)	\\
		$p.units$	& Units of pressure (bar)	\\
		$bin\_edges$	& Bin edges in wavenumbers (\cm) 	\\
		$bin\_edges.units$	& Units of bin edges (\cm) 	\\
		$xsecarr$	& Cross-section array ($p$, $t$, $bin\_centres$) 	\\
		$xsecarr.units$ &  Cross-section units, cm$^2$~/~molecule \\
		$mol\_mass$ & Molecular mass in a.m.u. \\
		$DOI$ & Digital Online Identifier for line list  \\
		$Date\_ID$ & ID for date of creation and version \\
		\hline\hline
	\end{tabular}
\end{table}

\subsection{ARCiS}\label{sec:arcis}

ARCiS is an atmospheric modelling and Bayesian retrieval package~\citep{20MiOrCh.arcis}.
Full details of the ARCiS code are presented in a separate paper,~\cite{20MiOrCh.arcis}. The most important information can be found in \cite{18OrMi.arcis}. The code consists of a forward modelling part based on correlated-k molecular opacities and cloud opacities using Mie and DHS \citep[Distribution of Hollow Spheres; see][]{05MiHoKo.arcis} computations. With ARCiS one can compute cloud formation~\citep{18OrMi.arcis} and chemistry \citep{18WoHeHu.exo} from physical and chemical principles. The code was  benchmarked against petitCODE \citep{15MoBoDu.petitCODE,17MoBoBo.exo} by \cite{18OrMi.arcis}. For the retrieval part the {\sc Multinest} algorithm \citep{08FeHo.multi,09FeGaHo.multi,13FeHoCa.multi} is employed. Benchmarks for the retrieval have been performed in the framework of the ARIEL mission~\citep{18PaBeBa.ARIEL}, showing excellent agreement with multiple other retrieval codes.
The Gauss sampling points used for the k-tables are based on Gauss Legendre polynomials. Opacities are in FITS format. 

\begin{table}[h]
	\caption{Overview of the data fields contained within the .fits k-tables for use in the ARCiS retrieval code. Here, $gauss$ refers to the gauss sampling points.}
	\label{t:ARCiS_tables} 
	\centering  
	\begin{tabular}{ll}
		\hline\hline
		\hline
		\rule{0pt}{3ex}Field name & Description	\\
		\hline\hline
		\rule{0pt}{3ex}$Tmin$	& Minimum temperature (K)  \\ 
		$Tmax$	& Maximum temperature (K)  \\ 
		$Pmin$	& Minimum pressure (bar)  \\ 
		$Pmax$	& Maximum pressure (bar)  \\ 
		$l\_min$	& Minimum wavelength ($\mu$m)  \\ 
		$l\_max$	& Maximum wavelength ($\mu$m)  \\ 
		$nT$	& Number of temperatures  \\ 
		$nP$	& Number of pressures  \\ 
		$nlam$	& Number of wavelength points  \\ 
		$ng$	& Number of gauss points  \\ 
		$kcoeff$ & k-coefficient array ($p$, $t$, $gauss$, $bin\_centre$)  \\ 
		$kcoeff\_units$ & k-coefficient units (cm$^2$~/~molecule) 	\\
		$t$	& List of temperatures (K)	\\
		$p$	& List of pressures (bar)	\\
		$bin\_centres$	& Bin centres in wavenumber (\cm) space	\\
		$mol\_mass$ & Molecular mass in a.m.u. \\
		$DOI$ & Digital Online Identifier for line list  \\
		$mol\_name$	& Molecule name \\
		$key\_iso\_ll$	& ID for isotopologue and line list	\\
		$Date\_ID$ & ID for date of creation and version \\
		\hline\hline
	\end{tabular}
\end{table}

\subsection{NEMESIS}\label{sec:nemesis}

NEMESIS is a planetary atmospheres radiative transfer and retrieval tool.
NEMESIS was originally developed for application to Solar System planets \citep[see, for example,][]{10TsWiBa.exo,11FlBaMo.exo} and has subsequently been extended and applied to exoplanets \citep[e.g.][]{12LeFlIr.exo,16BaAiIr.exo,18KrGaIr.exo,19IrPaTa.wasp43b}. It can be used with either an Optimal Estimation \citep{00Rodgers.exo} or Nested Sampling \citep[PyMultiNest;][]{ 08FeHo.multi,09FeGaHo.multi,13FeHoCa.multi,14BuGeNa.exo} algorithm. NEMESIS is capable of simulating a range of planetary radiative transfer scenarios, including exoplanet transit, eclipse and phase curve spectra, and nadir and limb sounding of Solar System atmospheres; it supports the inclusion of parameterised clouds, and for some geometries multiple scattering calculations can be performed.
Table~\ref{t:NEMESIS_ktables} gives a summary of the data contained within each NEMESIS~\citep{NEMESIS} k-table file, which are in binary format.

\begin{table}[h]
	\caption{Overview of the data fields contained within the binary k-tables for use in the NEMESIS retrieval code.}
	\label{t:NEMESIS_ktables} 
	\centering  
	\begin{tabular}{ll}
		\hline\hline
		\hline
		\rule{0pt}{3ex}Field name & Description	\\
		\hline\hline
		\rule{0pt}{3ex}$IREC0$	&  11 + 2*NG + 2 + NP + NT + NPOINT	\\
		$NPOINT$	& Number of wavelength points	\\
		$VMIN$	& Minimum wavelength ($\mu$m)	\\
		$DELV$	& 	-1.0 \\
		$FWHM$ & 0 \\
		$NP$	& Number of pressures 	\\
		$NT$	&  Number of temperatures	\\
		$NG$	&  Number of gauss points	\\
		$IDGAS1$	& NEMESIS ID for gas species	\\
		$ISOGAS1$	& NEMESIS ID for gas isotopologue \\
		$G\_ORDS$ & List of G-oordinates gauss points	\\
		$G\_WEIGHTS$ & List of weights for gauss points	\\
		$Blank$ &  Float for 0	\\
		$Blank$ &  Float for 0	\\
		$p$	& List of pressures (bar)	\\
		$t$	& List of temperatures (K)	\\
		$bin\_centres$	& Bin centres in wavelength ($\mu$m) space	\\
		$kcoeff$	& K-coefficient array ($\lambda$, $p$, $t$, $gauss$) \\
		& (10$^{20}$~cm$^2$~/~molecule) \\
		\hline\hline
	\end{tabular}
\end{table}

\subsection{petitRADTRANS}\label{sec:petit}

petitRADTRANS~\citep{19MoWaBo.petitRADTRANS} is an open-source radiative transfer code for exoplanet spectra with the Python package and implemented retrieval examples available on the code website\footnote{\url{https://petitradtrans.readthedocs.io/en/latest/}}. petitRADTRANS can calculate emission and transmission spectra for cloudy and cloud-free atmospheres, at high ($R=10^6$) and low ($R=1000$) resolution. The $R=1000$ branch of petitRADTRANS uses opacities in the form of k-tables, with a wavelength grid which differs slightly from that used in the k-tables and cross-sections produced for the other retrieval codes. The k-tables for petitRADTRANS are in HDF5 format, for similar reasons to those mentioned in Section~\ref{sec:taurex}. Table~\ref{t:petit_ktabs_hdf5} gives an overview of the data fields contained within the HDF5 k-tables for use in the petitRADTRANS retrieval code. The code available on the website has been updated to read the k-tables in HDF5 format, presented in this paper, directly. This works in a plug-and-play fashion. See the petitRADTRANS website for more information.

\begin{table}[h]
	\caption{Overview of the data fields contained within the HDF5 k-tables for use in the petitRADTRANS retrieval code.}
	\label{t:petit_ktabs_hdf5} 
	\centering  
	\begin{tabular}{ll}
		\hline\hline
		\hline
		\rule{0pt}{3ex}Field name & Description	\\
		\hline\hline
		\rule{0pt}{3ex}$bin\_centres$	& Bin centres in wavenumber (\cm) space	\\
		$bin\_edges$	& Bin edges in wavenumber (\cm) space	\\
		$wlrange$	& Wavelength ($\mu$m): min, max	\\
		$wnrange$	& Wavenumber (\cm): min, max	\\
		$samples$	& Gauss sampling points 	\\
		$weights$	& Weights for gauss sampling points 	\\
		$ngauss$	& Number of gauss sampling points 	\\
		$method$	& Description of sampling method 	\\
		$kcoeff$	& K-coefficient array ($p$, $t$, $bin\_centres$, $samples$) \\
		$kcoeff.units$ & K-coefficient units (cm$^2$~/~molecule) 	\\
		$t$	& List of temperatures	\\
		$p$	& List of pressures	\\
		$t.units$	& Units of temperature (K)	\\
		$p.units$	& Units of pressure (bar)	\\
		$mol\_mass$ & Molecular mass in a.m.u. \\
		$Date\_ID$ & ID for date of creation and version \\
		$DOI$ & Digital Online Identifier for line list  \\
		$mol\_name$	& Molecule name \\
		$key\_iso\_ll$	& ID for isotopologue and line list	\\
		\hline\hline
	\end{tabular}
\end{table}


\section{Line list sources and comments}\label{sec:sources}

Tables~\ref{t:sources_metal_oxides}~--~\ref{t:sources_ions} give details for all molecular data used in this work, with tables divided into groups, as on \url{www.exomol.com}. Included in each table is; the source for the line list used for each molecule (this is what is considered to be the most complete/accurate/up-to-date at the time of publication; the online ExoMol database will be updated to label the recommended line list for each molecular isotopologue if this changes), the associated minimum and maximum wavenumbers and wavelengths (E$_l$, E$_u$, $\lambda_l$, $\lambda_u$, respectively), number of lines in the line list, number of levels in the line list, the temperature up to which the line list is considered complete T$_{\max}$, and an indication for further comments, if applicable, which can be found in Section~\ref{sec:comments}.
Table~\ref{t:sources_atoms} gives the same information for select atomic species. 

		
	\begin{table}[H]
		\caption{Sources and properties of line list data used to compute the opacities presented in this work for metal oxides.}
		\label{t:sources_metal_oxides} 
		\centering  
		\begin{tabular}{lllcccccccccccccccccl}
			\hline\hline
			\rule{0pt}{3ex}Species	&	Line list	&	Ref	&	E$_l$ 	&	E$_u$ 	&	$\lambda_l$ 	&	$\lambda_u$ 	&	Lines	&	Levels	&	T$_{\max}$  & Notes	\\
				&		&		&	(\cm)	&	 (\cm)	&	 ($\mu$m)	&	 ($\mu$m)	&		&		&	 (K) & 	\\
			\hline
			\rule{0pt}{3ex}AlO	&	ExoMol ATP	&	\cite{jt598}	&	100	&	35,000	&	0.29	&	100	&	4.9 million	&	94,000	&	8000	&		\\
			CaO	&	ExoMol VBATHY	&	\cite{jt618}	&	100	&	20,000	&	0.5	&	100	&	28.4 million	&	130,000	&	5000	&		\\
			MgO	&	ExoMol LiTY	&	\cite{ExoMol_MgO}	&	100	&	33,000	&	0.3	&	100	&	72.8 million	&	190,000	&	5000	&		\\
			SiO	&	ExoMol EBJT	&	\cite{jt563}	&	100	&	6049	&	1.65	&	100	&	250,000	&	24,000	&	9000	&	(1a) 	\\
			TiO	&	ExoMol TOTO	&	\cite{ExoMol_TiO}	&	100	&	30,000	&	0.33	&	100	&	30 million	&	300,000	&	5000	&		\\
			VO	&	ExoMol VOMYT	&	\cite{jt644}	&	100	&	35,000	&	0.29	&	100	&	277 million	&	640,000	&	5000	&		\\
			\hline\hline
		\end{tabular}
	\end{table}

	\begin{table}[H]
		\caption{Sources and properties of line list data used to compute the opacities presented in this work for other oxides.  }
		\label{t:sources_other_oxides} 
		\centering  
		\begin{tabular}{lllcccccccccccccccccl}
			\hline\hline
			\rule{0pt}{3ex}Species	&	Line list	&	Ref	&	E$_l$ 	&	E$_u$  	&	$\lambda_l$ 	&	$\lambda_u$ 	&	Lines	&	Levels	&	T$_{\max}$  & Notes	\\
				&		&		&	(\cm)	&	 (\cm)	&	 ($\mu$m)	&	 ($\mu$m)	&		&		&	 (K) & 	\\
			\hline
			\rule{0pt}{3ex}CO	& Li 2015 		&	\cite{15LiGoRo.CO}	&	100	&	23,000	&	0.43	&	100	&	145,000	&	6400	&	5000 &		\\ 
			NO	&	 HITEMP-2019 	& \cite{jt763} 		&100		&	27,000	&	0.37	&	100	&	1.1 million	&	-	& 4000 & (1b)	\\ 
			O$_2$	&	HITRAN	&	\cite{HITRAN_2016}	&	100	&	6997	&  1.43		&	100	&	290,000	&	-	&	296	&	\\
			PO	&	ExoMol POPS	&	\cite{jt703}	&100		&	12,000	&	0.83	&	100	&	2.1 million	&	43,000	&	5000	&		\\
			\hline\hline
		\end{tabular}
	\end{table}
	
	\begin{table}[H]
		\caption{Sources and properties of line list data used to compute the opacities presented in this work for triatomics.  }
		\label{t:sources_triatomics} 
		\centering  
		\begin{tabular}{lllcccccccccccccccccl}
			\hline\hline
			\rule{0pt}{3ex}Species	&	Line list	&	Ref	&	E$_l$ 	&	E$_u$ 	&	$\lambda_l$ 	&	$\lambda_u$ 	&	Lines	&	Levels	&	T$_{\max}$  & Notes	\\
				&		&		&	(\cm)	&	 (\cm)	&	 ($\mu$m)	&	 ($\mu$m)	&		&		&	 (K) & 	\\
			\hline
			\rule{0pt}{3ex}CO$_2$	&	ExoMol UCL-4000	&	 \cite{20YuMeFr.co2}	& 100		&	20,000	&	0.5	& 100 &	8 billion	&	3.5 million	& 4000	&	(1c) 	\\
			H$_2$O	&	ExoMol 	&	\cite{jt734}	& 100		&	41,200	&	0.24	&	100	&	6 billion 	&	800,000	&	4000	& (1d)		\\
			&	 POKAZATEL	&&&&&&&	&&	\\
			H$_2$S	&	ExoMol AYT2	&	\cite{jt640}	&	100	& 11,000		&	0.91	&	100	&	115 million	&	220,000	&	2000	&		\\
			HCN	&	ExoMol Harris	&	\cite{jt570}	& 100		&	18,000	&	0.56	&	100	&	34.4 million	&	170,000	&	4000	&		\\
			N$_2$O	&	HITEMP-2019	&	\cite{jt763}	&	100 	&	12,900	&	0.76	&	100	& 3.6 million	&	-	&	1000	&		\\
			NO$_2$ &	HITEMP-2019 	& \cite{jt763}	 &	100 	&	4775	&	2.09	&	100	& 1.1 million	&	-	&	1000	&		\\
			O$_3$ &	HITRAN 	&	\cite{HITRAN_2016}	&	100	& 7000	&	1.43	&	100	&	290,000 &	-	&	296	&		\\
			SiH$_2$ & ExoMol CATS & \cite{20ClOwTe.sih2} & 100 & 10,000 & 1.00 &100 & 310 million &  594,000  & 2000 & \\
			SiO$_2$ & ExoMol OYT3 & \cite{20OwCoTe.sio2} &100 & 6000 &1.67  &100 & 32.9 billion & 5.7 million & 3000 & \\
			SO$_2$	&	ExoMol ExoAmes 	&	\cite{jt635}	&	100	&	8000	&	1.25	&	100	& 1.4 billion		&	3.3 million	&	2000	&		\\ 
			\hline\hline
		\end{tabular}
	\end{table}
	
	
	
	\begin{table}[H]
		\caption{Sources and properties of line list data used to compute the opacities presented in this work for metal hydrides.  }
		\label{t:sources_metal_hydrides} 
		\centering  
		\begin{tabular}{lllcccccccccccccccccl}
			\hline\hline
			\rule{0pt}{3ex}Species	&	Line list	&	Ref	&	E$_l$ 	&	E$_u$ 	&	$\lambda_l$ 	&	$\lambda_u$ 	&	Lines	&	Levels	&	T$_{\max}$  & Notes	\\
				&		&		&	(\cm)	&	 (\cm)	&	 ($\mu$m)	&	 ($\mu$m)	&		&		&	 (K) & 	\\
			\hline
			\rule{0pt}{3ex}AlH	&	ExoMol AlHambra	&	\cite{jt732}	&	100	& 27,000		&	0.37	&	100	& 36,000		&	1500	&	5000	&		\\
			BeH	&	ExoMol 	&	\cite{jt722}	&	100	&	42,000	&	0.24	&	100	&	590,000	&	15,000	&	2000	&		\\
			&	 Darby-Lewis	&&&&&&&	&&	\\
			CaH 	&  MoLLIST & \cite{11LiHaRa.CaH}		&	100	&	22,000	&	0.45	&	100	&	6000	&	914 	& 5000	&		\\
			CrH	&	MoLLIST	&	\cite{02BuRaBe.CrH}	& 100		&	14,500	&	0.69	& 100		&	13,800	&	1600	&	5000	& \\
			FeH	&	MoLLIST	&	\cite{10WEReSe.FeH}	&	100	&	15,000	&	0.67	&	100	&	93,000	&	3500	&	5000	&		\\
			LiH	&	CLT &	\cite{jt506} 	&	100	&	20,000	&	0.5	&	100	&	19,000	&	1100	&	2000	&		\\
			MgH 	&	MoLLIST	&	 \cite{13GhShBe.MgH}	&	100	&	29,000	&	0.34	&	100	&	14,200	&	1300	&	2000	& 	(1e)	\\
			NaH	& ExoMol Rivlin		&	\cite{jt605}	&	100	&	37,000	&	0.27	&	100	&	80,000	&	3300	&	7000	&		\\
			ScH	&	LYT	&	\cite{jt599} 	&	100	&	15,800	&	0.63	&	100	&	1.2 million	&	8500	&	2000	&		\\
			TiH	&	MoLLIST	&	\cite{05BuDuBa.TiH}	&	100	&	24,000	& 0.42		&	100	&	200,000	&	5800	&	5000	&		\\
			\hline\hline
		\end{tabular}
	\end{table}

	\begin{table}[H]
		\caption{Sources and properties of line list data used to compute the opacities presented in this work for other hydrides.  }
		\label{t:sources_other_hydrides} 
		\centering  
		\begin{tabular}{lllcccccccccccccccccl}
			\hline\hline
			\rule{0pt}{3ex}Species	&	Line list	&	Ref	&	E$_l$ 	&	E$_u$ 	&	$\lambda_l$ 	&	$\lambda_u$ 	&	Lines	&	Levels	&	T$_{\max}$  & Notes	\\
				&		&		&	(\cm)	&	 (\cm)	&	 ($\mu$m)	&	 ($\mu$m)	&		&		&	 (K) & 	\\
			\hline
			\rule{0pt}{3ex}CH	&	MoLLIST	& \cite{14MaPlVa.CH}	& 100		&	39,000	&	0.26	&	100	&	53,000	&	2500	&	5000	&		\\
			HBr & HITRAN &	\cite{13LiGoHa.hitran}	&	100	& 16,050	& 0.62		& 100		& 6070	&	-	&	5000	&		\\
			HCl	&	HITRAN 	&	\cite{13LiGoHa.hitran}	&	100	&	20,230	&	0.49	&	100	&	8890	&	-	&	5000	&		\\
			HF	&	HITRAN 	&	\cite{13LiGoHa.hitran}	&	100	&	32,350	& 0.31	&	100	&	8090	&	-	&	5000	&		\\
			HI &  HITRAN &	\cite{13LiGoHa.hitran}	&	100	& 14,000	&	0.71	&	100	& 3160	&	-	&	5000	& \\
			NH & MoLLIST & \cite{14BrBeWe.NH,15BrBeWe.NH} 	&	100	&	16,900	&	0.59	&	100	 	&	10,400	&	740 	&	5000	&		\\
			& & \cite{18FeBeHo.NH} 	&			&		&	 &		&		&		&		&		\\
			OH	& MoLLIST		&	\cite{16BrBeWe.OH} 	&	100 	&	43,400	&	0.23	&	100	&	54,000	&	1900	&	5000	&		\\
			& 		& \cite{18YoBeHo.OH}	&	 	&	&		&		&		&		&		&		\\
			PH	&	ExoMol LaTY	&	\cite{jt765}	&	100	&	24,500	&	0.41	&	100	&	64,800	&	2500	&	4000	&		\\
			SiH	&	ExoMol  SiGHTLY	&	\cite{jt711}	&	100	&	31,000	&	0.32	&	100	&	1.7 million	&	11,800	&	5000	&		\\
			SH	& ExoMol GYT &	\cite{jt776}	&	100	& 39,000	& 0.26	&	100	&	572,145 	&	7686	&	5000	&		\\
			\hline\hline
		\end{tabular}
	\end{table}
	
	
	
	\begin{table}[H]
		\caption{Sources and properties of line list data used to compute the opacities presented in this work for other diatomics.  }
		\label{t:sources_other_diatomics} 
		\centering  
		\begin{tabular}{lllcccccccccccccccccl}
			\hline\hline
			\rule{0pt}{3ex}Species	&	Line list	&	Ref	&	E$_l$ 	&	E$_u$ 	&	$\lambda_l$ 	&	$\lambda_u$ 	&	Lines	&	Levels	&	T$_{\max}$  & Notes	\\
				&		&		&	(\cm)	&	 (\cm)	&	 ($\mu$m)	&	 ($\mu$m)	&		&		&	 (K) & 	\\
			\hline
			\rule{0pt}{3ex}AlCl	&	MoLLIST	&	\cite{18YoBexx.AlF}	&	100	&	2350	&	4.26	&	100	&	20,200	&	2400	& 5000		&		\\
			AlF	&	MoLLIST	&	\cite{18YoBexx.AlF}	&	100	&	3880	&	2.58	&	100	&	40,500	&	2420	&	5000	&		\\ 
			C$_2$	&	 ExoMol 8states	&	\cite{18YuSzPy.C2}	&	100	&	48,660	&	0.21	&	100	&	6.1 million	&	44,190	&	5000	&		\\
			CaF	&	MoLLIST	&	\cite{18HoBexx.CaF}	&	100	&	5580	&	1.79	&	100	&	14,800	&	1360	&	5000	&		\\  
			CN	&	MoLLIST	&	\cite{14BrRaWe.CN}	&	100	&	44,200	&	0.23	&	100	&	195,000	& 7700		&	5000	&		\\
			CP	&	MoLLIST	&	\cite{14RaBrWe.CP}	&	100	&15,000		&	0.67	&	100	&	28,700	&	2100	&	5000	&		\\
			CS	&	ExoMol JnK	&	\cite{jt615}	&	100	&	11,000	&	0.91	&	100	&	199,000	&	11,500	&	3000	&		\\ 
			H$_2$	& RACPPK &	\cite{19RoAbCz.H2} 	&	100	&	36,000	&	0.28	&	100	&	4700	&	300	&	5000	&		\\
			KCl	&	ExoMol Barton	&	\cite{jt583}	&	100	&	2900 &	3.45	& 100	&	1.3 million	&	60,700	&	3000	&		\\
			KF	&	MoLLIST	&	\cite{16FrBeBr}	&	100	&	4000	&	2.49	&	100	&	10,500	&	1060	&	5000	&		\\
			LiCl	&	MoLLIST	&	\cite{18BiBexx}	&	100	&	4840	&	2.07	&	100	&	26,200	&	2400	&	5000	&		\\
			LiF	&	MoLLIST	&	\cite{18BiBexx}	&	100	&	1810	&	5.52	&	100	&	10,600	&	2400	&	5000	&		\\ 
			MgF	&	MoLLIST	&	\cite{17HoBexx.MgF}	&	100	&	5470	&	1.83	&	100	&	8100	&	900	&	5000	&		\\ 
			NaCl	&	ExoMol Barton	&	\cite{jt583}	&	100	&	2500	&	4.00	& 100	&	703,000	& 49,000		&	3000 	&		\\
			NaF	&	 MoLLIST	&	\cite{16FrBeBr}	&100		&	4990	&	2.01	&	100	&	7900	&	840	&	5000	&		\\ 
			NS	&	ExoMol SNaSH	&	\cite{jt725}	& 100		& 38,420		&	0.26	&	100	&	3.2 million &	31,500	&	5000	&		\\
			PN	& ExoMol YYLT	&	\cite{14YoYuLo.PN}	& 100		&	6500	&	1.54	&	100	&	140,000	&	14,000	&	5000	&		\\ 
			PS	&	ExoMol POPS	&	\cite{jt703}	& 100		&	36,700	& 0.27		&	100	& 30.4 million	&	226,000	&	5000	&		\\ 
			SiS	&	ExoMol UCTY	&	\cite{jt724}	&	100	&	3700	&	2.70	&	100	&	91,600	&	10,000	&	5000	&		\\ 
			\hline\hline
		\end{tabular}
	\end{table}
	
	
	\begin{table}[H]
		\caption{Sources and properties of line list data used to compute the opacities presented in this work for larger molecules.  }
		\label{t:sources_larger_molecules} 
		\centering  
		\begin{tabular}{lllcccccccccccccccccl}
			\hline\hline
			\rule{0pt}{3ex}Species	&	Line list	&	Ref	& E$_l$ 	&	E$_u$  	&	$\lambda_l$ 	&	$\lambda_u$ 	&	Lines	&	Levels	&	T$_{\max}$  & Notes	\\
				&		&		&	(\cm)	&	 (\cm)	&	 ($\mu$m)	&	 ($\mu$m)	&		&		&	 (K) & 	\\
			\hline
			\rule{0pt}{3ex}AsH$_3$	&	ExoMol CYT18	&	\cite{jt751}	&	100	&	7000	&	1.43	&	100	&	3.6 million	&	4.3 million	  &	296	&		\\ 
			C$_2$H$_2$	&	ExoMol aCeTY	&	\cite{jt780}	&	100	&	10,000	&	1.00	&	100	&	4.3 billion	&	5.2 million	&	2200	&	(1f)	\\
			C$_2$H$_4$	& ExoMol MaYTY		&	\cite{jt729}	&	100	&7100		&1.41		&	100	&	50 billion	&	45 million	& 700	&	\\ 
			CH$_3$ &	ExoMol AYYJ	&	\cite{19AdYaYu.CH3}	&	100	&	10,000	&	1.00	&	100		&	2.1 billion	&	9.1 million	&	1500	&		\\
			CH$_3$Cl	&	ExoMol OYT	&	\cite{jt733}	&	100	&	6400	&	1.56	&	100	&	166 billion	&	10.2 million	&	1200	&		\\
			CH$_3$F	&	ExoMol OYKYT	&	\cite{19OwYaKu}	&	100	&	4700	&2.13		&	100	&	1.4 billion	&	3.5 million	& 300		&		\\ 
			CH$_4$	&	ExoMol 34to10	&	\cite{jt698}	&	100	&	18,000	& 0.56		& 100 & 34 billion	&	8.2 million 	&	2000	&	(1g)	\\ 
			H$_2$O$_2$	&	ExoMol APTY	&	\cite{jt638}	&	100	&	6000	&	1.67	&	100	&	10 billion	&	7.6 million	&	1250	&		\\ 
			H$_2$CO	&	ExoMol AYTY	&	\cite{jt597}	&100		&10,100		&	0.99	&	100	&	10 billion	&	10.3 million	&	1500	&		\\ 
			HNO$_3$	&	ExoMol AIJS	&	\cite{15PaYuTe.exo}	&	100	&	7100	& 1.41		&	100	&	7 billion	&	17.5 million	&	500	&		\\ 
			NH$_3$	&	ExoMol CoYuTe	&	\cite{jt771}	&	100	&	20,000	&	0.5	&	100	&	16.9 billion	&	5.1 million	&	1500	&		\\
			P$_2$H$_2$ 	&	OY-Cis	&	\cite{19OwYuxx.P2H2}	&	100	&	6000	&	1.67	&	100	&	5.9 billion	&	6 million	&	300	&		\\ 
			(cis)&	 	&&&&&&&	&&	\\
			P$_2$H$_2$ 	&	OY-Trans	&	\cite{19OwYuxx.P2H2}	&		100	&	6000	&	1.67	&	100	&	5.3 billion	& 5.9 million	&	300	&		\\
			(trans)&	 	&&&&&&&	&&	\\
			PH$_3$ &	ExoMol SAlTY	&	\cite{jt592}	&	100	&	10,000	&	1.00	&	100	&	16.8 billion	& 9.8 million	&	1500	&		\\
			SiH$_4$	&	ExoMol OY2T	&	\cite{jt701}	&	100	&	5000	&	2.00	&	100		&	62.7 billion	&	7.1 million	&	1200	&		\\
			SO$_3$	&	ExoMol UYT2	&	\cite{jt641}	&	100	&	5000	&	2.00	&	100	&	21 billion	& 18.5 million	&	800	&		\\ 
			\hline\hline
		\end{tabular}
	\end{table}
	
	\begin{table}[H]
		\caption{Sources and properties of line list data used to compute the opacities presented in this work for ions.  }
		\label{t:sources_ions} 
		\centering  
		\begin{tabular}{lllcccccccccccccccccl}
			\hline\hline
			\rule{0pt}{3ex}Species	&	Line list	&	Ref	&	E$_l$ 	&	E$_u$  	&	$\lambda_l$ 	&	$\lambda_u$ 	&	Lines	&	Levels	&	T$_{\max}$  & Notes	\\
				&		&		&	(\cm)	&	 (\cm)	&	 ($\mu$m)	&	 ($\mu$m)	&		&		&	 (K) & 	\\
			\hline
			\rule{0pt}{3ex}H$_3$$^{+}$	&	ExoMol MiZATeP	&	\cite{jt666}	&100		&25,000		&	0.4	&	100	&	127.5 million	&	159,000	&	5000	&		\\
			H$_3$O$^{+}$	&	ExoMol eXeL &	\cite{20YuTeMi.h3oplus}	&	100	&	10,000	&	1.00	&	100	&	2.1 billion	&	1.2 million	&	1500	&		\\
			HD$^{+}$	&	ADJSAAM	 	&	\cite{19AmDiJo.HeHplus}	&	100	&	21,500	&	0.47	&	100	&	10,300	&	640	& 4000	&	(1h)		\\
			HeH$^{+}$ &	ADJSAAM	&	\cite{19AmDiJo.HeHplus}	&100		&	14,900	&	0.67	&	100	&	1400	&	180	&	4000 &	(1i)		\\ 
			LiH$^{+}$	&	CLT	&	\cite{jt506}	& 100		&	920	&	10.87	&	100	&	330	&	75	&	2000	&		\\ 
			OH$^{+}$	&	MoLLIST	&	\cite{17HoBexx.OH+}	&	100	&	30,300	&	0.33	&	100	&12,000		&	820	&	5000	&		\\ 
			\hline\hline
		\end{tabular}
	\end{table}
	\begin{table}[H]
		\caption{Sources and properties of line list data used to compute the opacities presented in this work for atoms.  }
		\label{t:sources_atoms} 
		\centering  
		\begin{tabular}{lllcccccccccccccccccl}
			\hline\hline
			\rule{0pt}{3ex}Species	&	Line list	&	Ref	& E$_l$ 	&	E$_u$ 	&	$\lambda_l$ 	&	$\lambda_u$ 	&	Lines	&	Levels	&	T$_{\max}$  & Notes	\\
				&		&		&	(\cm)	&	 (\cm)	&	 ($\mu$m)	&	 ($\mu$m)	&		&		&	 (K) & 	\\
			\hline
			\rule{0pt}{3ex}K	&	NIST	&	\cite{NISTWebsite}	&	100	&	35,000	&	0.29	&	100	&	186	&	188	&	5000	&	(1j)	\\
			Na	&	NIST	&	\cite{NISTWebsite}	&	100	&	42,000	&	0.24	&	100	&	523	&	117	&	5000	& (1j)		\\ 
			\hline\hline
		\end{tabular}
	\end{table}




\subsection{Comments on tables}\label{sec:comments_tables}

\subsubsection{General comments}

\begin{itemize}
	\item The temperature of completeness for all the MoLLIST~\citep{MOLLIST} molecules is assumed to be 5000~K, although it is expected that most of these will not necessarily be complete, even at lower temperatures. Nevertheless, they are the best data currently available for these molecules. The opacities for these species will be updated in the future, if and when new line list data is produced.
	\item There are data available for some molecules not mentioned in the Tables of this work, such as PF$_3$~\citep{jt752}, which is currently only computed up to a low value of the rotational angular momentum quantum number, $J$. It is therefore incomplete in its current state, but can be computed if requested. Readers are encouraged to contact the ExoMol team if there are particular requests for molecular data not already computed. 
	\item As previously mentioned, a small number of molecules in the HITRAN~\citep{HITRAN_2016} database (HF, HCl, HBr, HI, H$_2$) are considered applicable up to high temperatures of around 4000~-5000~K~\citep{13LiGoHa.hitran}. 
\end{itemize}


\subsubsection{Comments on individual species}\label{sec:comments}
\noindent
\textbf{(1a)} Work is underway for an updated ExoMol line list for SiO which will extend in the ultraviolet. The current line list only considers vibration-rotation transitions and so the current maximum wavenumber was set at 6049~\cm. \\
\textbf{(1b)} The HITEMP line list for NO includes data from the ExoMol NOname line list \citep{jt686}. \\
\textbf{(1c)} The Ames line list~\citep{17HuScFr.co2} and the CDSD-4000 databank~\citep{11TaPe.co2} are also available for CO$_2$, as well as the HITEMP compilation~\citep{HITEMP}. \\
\textbf{(1d)} The previous ExoMol line list for H$_2$O, BT2~\citep{jt378} is only complete up to temperatures of 3000~K, whereas the more accurate ExoMol POKAZATEL line list~\cite{jt734} is complete up to 4000~K. \\
\textbf{(1e)} There is also a line list for MgH from ExoMol Yadin~\citep{jt529}. However, since it only covers the ground electronic $X^2\Sigma^{+}$ state, and so is less complete than the more recent MoLLIST line list of \cite{13GhShBe.MgH}, we use the latter.\\
\textbf{(1f)} Previous to the ExoMol aCeTY line list of \cite{jt780}, the main sources of data for acetylene were from HITRAN~\citep{HITRAN_2016} and ASD-1000~\cite{17LyPe.C2H2}. The data from HITRAN is only applicable for room temperature studies, and was shown in \cite{jt780} to be inadequate for high-temperature applications. ASD-1000 was a vast improvement, although there does seem to be opacity missing from some of the hot bands when compared to ExoMol aCeTY in \cite{jt780}. \\
\textbf{(1g)} The previous ExoMol line list for CH$_4$, called 10to10 \cite{jt564}, is only complete up to 1500~K. The updated 34to10 line list is therefore recommended instead. Future updates of the database will investigate using data for methane based on recent line lists from either TheoReTs \citep{17ReNiTyi} or HITEMP \cite{20HaGoRe.CH4}; these are currently expected to be more accurate when considering high-resolution applications. For low-resolution applications, we expect the quality of the ExoMol line list used here to be sufficient, particularly because completeness is more important than accuracy at lower resolutions \citep{jt572}. \\
\textbf{(1h)} The energy states from \cite{jt506} are used in the \cite{19AmDiJo.HeHplus} line list for HD$^{+}$.  \\
\textbf{(1i)} The energy states from \cite{jt347} are used in the \cite{19AmDiJo.HeHplus} line list for HeH$^+$. \\
\textbf{(1j)} The pressure and temperature broadened profiles for the resonance doublets of Na and K are computed using \cite{16AlSpKi.broad} and \cite{19AlSpLe.broad}. See section \ref{sec:atomic_broad} for a discussion on the broadening profiles of these atoms. \\


\subsection{Isotopologues}\label{sec:isos}

For the majority of species, we provide the opacities for the main isotopologue only, or separate opacity files for the other isotopologues. For some species, however, it is important to take natural abundances into account \citep[see, for example,][]{NISTWebsiteIsos}. 
We therefore provide opacity files combined at natural abundances for the species listed in Table~\ref{t:isos}, as well as the separate isotopologue opacities. 


\begin{table}[h]
	\caption{Species where the opacities combined according to natural elemental abundances are provided.
		 NA is the natural abundance, and MM is the molecular mass of each isotopologue.}
	\label{t:isos} 
	\begin{tabular}{llll}
		\hline\hline
		\hline
		\rule{0pt}{3ex}Species 	&	Iso	&	NA (\%)	&	MM (a.m.u.)	\\
		\hline\hline
		\rule{0pt}{3ex}TiO 	&	$^{48}$Ti$^{16}$O	&	73.7	&	63.94		\\
		TiO 	&	$^{46}$Ti$^{16}$O	&	8.3	&	61.94		\\
		TiO 	&	$^{47}$Ti$^{16}$O	&	7.4	&	62.94		\\
		TiO 	&	$^{49}$Ti$^{16}$O	& 5.4	&	64.94		\\
		TiO 	&	$^{50}$Ti$^{16}$O	&	5.2	&	65.94		\\
		\hline
		\rule{0pt}{3ex}CO	&	$^{12}$C$^{16}$O	&	98.7	&	27.99		\\
		CO	&	$^{13}$C$^{16}$O	&	1.1	& 28.99
		\\
		CO	&	$^{12}$C$^{18}$O	&	2.0~$\times$~10$^{-3}$	& 29.99
		\\
		CO	&	$^{12}$C$^{17}$O	&	3.7~$\times$~10$^{-4}$	& 28.99
		\\
		CO	&	$^{13}$C$^{18}$O	&	2.2~$\times$~10$^{-5}$	& 31.00
		\\
		CO	&	$^{13}$C$^{17}$O	&	4.1~$\times$~10$^{-6}$	& 30.00 \\
		\hline
		\rule{0pt}{3ex}HBr	&	H$^{79}$Br	&	50.7	&	79.93		\\
		HBr	&	H$^{81}$Br	&	49.3	& 81.92 \\
		HBr	&	D$^{79}$Br	&	7.9~$\times$~10$^{-3}$	&	80.93		\\
		HBr	&   D$^{81}$Br	&	7.7~$\times$~10$^{-3}$	& 82.93  \\
		\hline
		\rule{0pt}{3ex}HCl	&	H$^{35}$Cl	&	75.8	&		35.98	\\
		HCl	&	H$^{37}$Cl	&	24.2	&  37.97 \\
		HCl	&	D$^{35}$Cl	&	1.2~$\times$~10$^{-2}$	&	36.98		\\
		HCl	&	D$^{37}$Cl	&	3.8~$\times$~10$^{-3}$	&  38.98  \\
				\hline
		\rule{0pt}{3ex}CH$_3$Cl	&	$^{12}$CH$_3$$^{35}$Cl	&	74.9	&	49.99		\\
		CH$_3$Cl	&	$^{12}$CH$_3$$^{37}$Cl	&	23.9	&	51.99  \\
						\hline
		\rule{0pt}{3ex}KCl	&	$^{39}$K$^{35}$Cl	&  70.6 &73.93	\\
		KCl	&	$^{39}$K$^{37}$Cl	& 22.6 & 75.93 \\
		KCl	&	$^{41}$K$^{35}$Cl	&5.1 &	75.93 \\
		KCl	&	$^{41}$K$^{37}$Cl	& 1.6 & 77.92 \\
			\hline
		\rule{0pt}{3ex}NaCl	&	$^{23}$Na$^{35}$Cl	& 75.8	&57.96	\\
		NaCl	&	$^{23}$Na$^{37}$Cl	&	24.2	& 59.96 \\
			\hline
		\rule{0pt}{3ex}LiCl	&	$^{7}$Li$^{35}$Cl	& 70.0 & 41.98	\\
		LiCl	&	$^{7}$Li$^{37}$Cl	& 22.4  & 43.98 \\
		LiCl	&	$^{6}$Li$^{35}$Cl	& 5.8 & 40.98 \\
		LiCl	&	$^{6}$Li$^{37}$Cl	&  1.8 & 42.98 \\
					\hline
		\rule{0pt}{3ex}AlCl	&	$^{27}$Al$^{35}$Cl	& 75.8 & 61.95	\\
		AlCl	&	$^{27}$Al$^{37}$Cl	& 24.2	& 63.95 \\
					\hline
		\rule{0pt}{3ex}MgO	&	$^{24}$Mg$^{16}$O	& 78.8 & 39.98	\\
		MgO	&	$^{25}$Mg$^{16}$O	& 9.9 & 40.98 \\
		MgO	&	$^{26}$Mg$^{16}$O	& 10.9 & 41.98 \\
		MgO	&	$^{24}$Mg$^{17}$O	& 3.0~$\times$~10$^{-2}$ &  40.98 \\
		MgO	&	$^{24}$Mg$^{18}$O	& 0.2 & 41.98 \\
							\hline
		\rule{0pt}{3ex}MgH	&	$^{24}$MgH	& 79.0 &  24.99	\\
		MgH	&	$^{25}$MgH	& 10.0 & 25.99 \\
		MgH	&	$^{26}$MgH	&11.0 & 26.99 \\
							\hline
		\rule{0pt}{3ex}MgF	&	$^{24}$Mg$^{19}$F	& 79.0  & 42.98	\\
		MgF	&	$^{25}$Mg$^{19}$F	& 10.0 & 43.98 \\
		MgF	&	$^{26}$Mg$^{19}$F	& 11.0 & 44.98 \\
		\hline\hline
	\end{tabular}
\end{table}

The species included in this Table may be updated if new isotopologue line lists become available \citep[line lists can be extended to different isotopologues using the method outlined in][with line positions approaching experimental accuracy for species such as H$_2$O]{jt665}. A summary of the number of isotopologues available for various line lists can be found in \cite{18TeYu.exo}.

\section{Visible and UV}\label{sec:vis_UV}

The wavelength regions covered by the data for each molecule should be carefully noted (seeI ables~\ref{t:sources_metal_oxides}~--~\ref{t:sources_ions}); many of the diatomic molecules (such as PO, SiO, CrH, FeH, NH, PN, KCl, NaCl, LiCl, CS, CP, AlCl, AlF, KF, LiF, CaF, MgF) are not  covered for
wavelengths short of 0.67~$\mu$m, even though they are expected to have opacity in this region. Using our opacities for such species to characterise observations which span beyond this region is therefore not advised, as it could give the impression that the opacity suddenly drops off at the wavelength at which the data ends, which in practise would not be true. It is therefore desirable to have these opacities computed to higher energies (corresponding to lower wavelengths). At present some data are available in these regions for certain molecules. The ExoMol
project is planning further work on this problem which will need to consider
bound - free (photodissociation) as well as bound-bound processes. 



\section{The ExoMolOP database}\label{sec:exomolOP}

The ExoMolOP database comprises opacity data for over 80 species, details of which can be found in 
Tables~\ref{t:sources_metal_oxides}~--~\ref{t:sources_atoms}. The data are formatted in different ways for four different exoplanet atmosphere retrieval codes; ARCiS, TauREx, NEMESIS and petitRADTRANS, (see Section~\ref{sec:ret_codes}) and include cross-sections (at R~=~$\frac{\lambda}{\Delta \lambda}$~=~15,000) and k-tables (at R~=~1000) for the 0.3~-~50$\mu$m wavelength region. Voigt profiles are used to represent the broadening of molecular lines, using the broadening parameters detailed in Tables~\ref{t:ch4}~-~\ref{t:hcn}, with line wings computed to 500~Voigt widths from the line centres. The pressure and temperature broadened profiles for the atomic resonance doublets of Na and K are computed using the tables of \cite{16AlSpKi.broad} and \cite{19AlSpLe.broad}.

\subsection{Opacity Data Location}\label{sec:location}

The opacity database is available at \url{www.exomol.com}. Opacity files can be downloaded from \url{www.exomol.com/data/data-types/opacity/} and used directly for four retrieval codes; ARCiS, TauREx, NEMESIS and petitRADTRANS, but are intended to be sufficiently easy to manipulate for general use also. 

The data will be fully integrated into ExoMol and will form part of the 2020 release which has just been completed \citep{jt804}. The opacity cross sections and $k$-tables will also be made available via the virtual atomic and molecular data centre (VAMDC) portal
\citep{jt481,jt630}.

\subsection{Keeping opacity data up-to-date}

The ExoMol application programming interface (API) is described in \cite{jt631}, with an associated master definition file, ExoMol.all, available from \url{www.exomol.com/exomol.all}. The "def" files are accessed at URLs of the form \url{www.exomol.com/db/<molecule>/<iso-slug>/<dataset-name>/<iso-slug>__<dataset-name>.def}, with the latest dataset-name for a particular molecule given in ExoMol.all. This allows for automated updates of ExoMol data to be converted to opacity (cross-section and k-table) data, using the HTTP WGET formalism. We use the same format as described in the ExoMol master definition files for our cross-section and k-table file naming (also $key\_iso\_ll$ within the files; see Section~\ref{sec:ret_codes}). Opacity files can therefore be downloaded using a URL in the following form for each of the four retrieval code:\\
\url{www.exomol.com/db/<molecule>/<iso-slug>/<dataset-name>/<iso-slug>__<dataset-name>.R1000_0.3-50mu.ktable.NEMESIS.kta} \\
\url{www.exomol.com/db/<molecule>/<iso-slug>/<dataset-name>/<iso-slug>__<dataset-name>.R15000_0.3-50mu.xsec.TauREx.h5} \\
\url{www.exomol.com/db/<molecule>/<iso-slug>/<dataset-name>/<iso-slug>__<dataset-name>.R1000_0.3-50mu.ktable.ARCiS.fits.gz } \\
\url{www.exomol.com/db/<molecule>/<iso-slug>/<dataset-name>/<iso-slug>__<dataset-name>.R1000_0.3-50mu.ktable.petitRADTRANS.h5} \\
 Most of the files produced have a version ID field contained within ($Date\_ID$), so future updates can be tracked. For example, if the cross-sections or k-tables are recomputed with improved broadening parameters, new versions will be released. It should also be noted that studies which investigate the computational feasibility vs retrieval accuracy of both k-tables and cross-sections were made before some more recent computational improvements, such as more widespread use of GPUs \citep[see, for example,][]{19AlChWa.taurex}, and similar investigations may be beneficial. Future datasets may therefore be computed at higher R~=~$\frac{\lambda}{\Delta \lambda}$ to reflect requirements. It should also be noted data from high-resolution observations requires line-by-line integrated opacities for analysis, as discussed below, which are typically only computed in the wavelength region necessary to match observations. Such opacities are not included in the present database, but may be added in the future. For these reasons, users of the ExoMolOP database are strongly advised to reference the version of the opacities used in publications, along with the associated line list.



\section{High-resolution opacity requirements}\label{sec:highres}

It is stressed that this database is not intended for high-resolution applications. It would be beneficial to compute a series of very high resolution cross-sections, but restricted only to wavelength regions necessary to match available observational data for cross-correlation studies; see, for example, \cite{14KoBiBr.exo,18HaMaCa.exo,19MoSn.exo,20WeBrGa.exo}. Currently, only a small sample of the line lists which are detailed in Tables~\ref{t:sources_metal_oxides}~--~\ref{t:sources_ions} of Section~\ref{sec:sources} are suitable for use in high-resolution applications, with cross-sections typically required to be sampled to a resolution of at least R~=~$\frac{\lambda}{\Delta \lambda}$~=~100,000. This includes those molecules which are found in the HITEMP~\citep{HITEMP}, HITRAN~\cite{HITRAN_2016}, or MoLLIST~\citep{MOLLIST} databases, and only those in the ExoMol~\citep{jt528,jt631} database which have been ``MARVELised''. MARVEL (measured active vibration-rotation energy levels) is a prodecure whereby transition wavenumbers from all available laboratory experiments are analysed together to produce a list of experimentally-determined energy levels \citep{jt412,jt750}. These empirical energies  are then subsequently included in the ``MARVELised'' ExoMol line lists to improve their accuracy (see, for example, \cite{jt705,jt780,20McSyBo.C2}). As only a sub-section of the energy levels, and therefore transitions, which are included in an ExoMol line list will have been MARVELised, the new ExoMol data format includes an uncertainty column in the .states file~\citep{jt804}. This gives an indication of how reliable an individual energy level is, and therefore all transitions which involve this level. 
A set of high-resolution cross-sections for six molecular species has recently been made publicly available by \cite{jt782}.

\section{Conclusion}\label{sec:conclusion}

In this work we present a publicly available database of opacity cross-sections and k-tables for molecules of astrophysical interest, ExoMolOP, which is primarily based on the latest line list data from the ExoMol~\citep{jt528,jt631}, HITEMP~\citep{HITEMP,jt763} and MoLLIST~\citep{MOLLIST} databases. These data are generally suitable for characterising high temperature exoplanet or cool stellar atmospheres, and have been computed at a variety of pressures and temperatures, with a few molecules included at room-temperature only from the HITRAN database. The data are formatted in different ways for four different exoplanet atmosphere retrieval codes; ARCiS~\citep{20MiOrCh.arcis}, TauREx~\citep{15WaTiRo.taurex,15WaRoTi.taurex,19AlChWa.taurex}, NEMESIS~\citep{NEMESIS}, and petitRADTRANS~\citep{19MoWaBo.petitRADTRANS}. Opacity data for Na and K are also included using line list data from the NIST~\citep{NISTWebsite} database and the broadening scheme of \cite{16AlSpKi.broad,19AlSpLe.broad}. 

New opacities will be added to the ExoMolOP database as new or updated line lists become available. 
Updating an existing line list may include the ``MARVEL'' procedure (which can lead to noticeable differences in line positions, even at low resolution). As previously mentioned, the current ExoMolOP database is not suitable for high-resolution cross-correlation studies, but high-resolution opacities may be provided in the future alongside the current database. There are some opacities for high-resolution Doppler shift studies currently available from \cite{jt782}. 
 For the current release of the ExoMolOP database, a number of assumptions related to the broadening parameters were made, which can clearly be improved upon. The upcoming ExoMolHD project will focus on this explicitly for the next release of the database. For that, use will be made of advances in molecular broadening and line-shape theory~\citep{20StThCy.broad,18HaTrAr.broad}.
  Even though the broadening parameters used in this work can clearly be improved, they are generally considered adequate for low-resolution retrieval studies. Particular care should be taken, however, when using these opacities in regimes where the broadening parameters may have more effect, such as for low pressure and low temperature environments.  

 Currently, only H$_2$ and He parameters have been used, as the primary intention is for the characterisation of ``Hot Jupiters'' exoplanets. However, future releases of the database will extend to other types of broadener, such as self-broadening, N$_2$, CO$_2$ and air, which are thought to be important in other types of planet, such as ``mini-Neptunes'' or ``Super Earths''.  It is thought that, for some molecules in particular, for example H$_2$O, the differences between self-broadening parameters can be around 7~times larger than the H$_2$~/~He parameters~\citep{05BrBeMa.broad,16PtMcPo.broad,19GhLi.broad}. 


There are often different versions of computed line lists for one molecule, some with quite stark differences. Users of the ExoMolOP database are strongly urged to include citations to the line list relating to the opacities used in their publications. We include a bibtex file with the citations to all opacities included as part of the supplementary data to this work. 





\section{Acknowledgements}

This project has received funding from the European Union's Horizon 2020 Research and Innovation Programme, under Grant Agreement 776403, and from the European Research Council (ERC) under the European Union's Horizon 2020 research and innovation programme under grant agreement No 758892, ExoAI. P.M. acknowledges support from the European Research Council under the European Union's Horizon 2020 research and innovation program under grant agreement No. 832428. J.T. and S.Y. thank the STFC Project No. ST/R000476/1. 

We gratefully acknowledge Cambridge Service for Data Driven Discovery (CSD3), part of which is operated by the University of Cambridge Research Computing on behalf of the STFC DiRAC HPC Facility (\url{www.dirac.ac.uk}). The DiRAC component of CSD3 was funded by BEIS capital funding via STFC capital grants ST/P002307/1 and ST/R002452/1 and STFC operations grant ST/R00689X/1. DiRAC is part of the National e-Infrastructure.

We would like to thank Nicole Allard for providing all the data necessary to compute the broadening profiles for the Na and K resonance doublets. We thank the reviewer for their comments to improve the manuscript. 

\bibliographystyle{aa}
\bibliography{ExoMolOP}

\onecolumn
\begin{appendix}
	\section{Broadening tables}

	\begin{table}[H]
		\caption{Molecular properties for the nonpolar species with computed opacities presented in this work, where insufficient broadening parameters were found in the literature, and so those of CH$_4$ (see Table~\ref{t:sources_2}) were used instead. DM is the dipole moment, MM is the molar mass, and AN is the atomic number.}
		\label{t:ch4} 
		\centering  
		\begin{tabular}{llllll}
			\hline\hline
			\hline
			\rule{0pt}{3ex}Species 	&	DM	&	MM	&	AN	& Structure &	Dipole Ref	\\
			&	(d)	&	(g/mol)	&		&		\\
			\hline\hline
			\rule{0pt}{3ex}SiH$_4$	&	0 	&	32.1	&	18	&Nonpolar	&	\\
		\end{tabular}
	\end{table}
	
	\begin{table}[H]
		\caption{Molecular properties for the nonpolar species with computed opacities presented in this work, where insufficient broadening parameters were found in the literature, and so those of H$_2$ (see Table~\ref{t:sources_2}) were used instead. DM is the dipole moment, MM is the molar mass, and AN is the atomic number.}
		\label{t:h2} 
		\centering  
		\begin{tabular}{llllll}
			\hline\hline
			\hline
			\rule{0pt}{3ex}Species 	&	DM	&	MM	&	AN	& Structure &	Dipole Ref	\\
			&	(d)	&	(g/mol)	&		&		\\
			\hline\hline
			\rule{0pt}{3ex}C$_2$	&	0	&	12.0	&	12	&	Diatomic	&	\cite{CCCBDBWebsite}	\\
			H$_3$$^{+}$	&	0	&	3.0	&	3	&	Nonpolar	&	\cite{CCCBDBWebsite}	\\
			HD$^{+}$	&	0.1	&	3.0	&	2	&	Diatomic	&	\cite{74Bunker.hdplus}	\\
		\end{tabular}
	\end{table}

	\begin{table}[H]
		\caption{Molecular properties for the nonpolar species with computed opacities presented in this work, where insufficient broadening parameters were found in the literature, and so those of C$_2$H$_2$ (see Table~\ref{t:sources_2}) were used instead. DM is the dipole moment, MM is the molar mass, and AN is the atomic number.}
		\label{t:c2h2} 
		\centering  
		\begin{tabular}{llllll}
			\hline\hline
			\hline
			\rule{0pt}{3ex}Species 	&	DM	&	MM	&	AN	& Structure &	Dipole Ref	\\
			&	(d)	&	(g/mol)	&		&		\\
			\hline\hline
			\rule{0pt}{3ex}C$_2$H$_4$	&	0	&	28.0	&	16	&	Nonpolar	&		\\
			CH$_3$ &	0 	&	15.0	&	9	& Nonpolar  &	\cite{CCCBDBWebsite}	\\
			CO$_2$	&	0	&	44.0	&	22	&	Linear	&		\\
			P$_2$H$_2$ (trans)	&	0	&	64.0	&	32	&	Nonpolar	&		\\
			SiO$_2$ & 0 & 60.1 &  30 & Linear & \\
			SO$_3$	&	0 	&	80.1	&	40	& Nonpolar	&	\\
			\hline\hline
		\end{tabular}
	\end{table}

	\begin{table}[H]
		\caption{Molecular properties for the species with computed opacities presented in this work, where insufficient broadening parameters were found in the literature, and so those of CO (see Table~\ref{t:sources_2}) were used instead. DM is the dipole moment, MM is the molar mass, and AN is the atomic number.}
		\label{t:co} 
		\centering  
		\begin{tabular}{llllll}
			\hline\hline
			\hline
			\rule{0pt}{3ex}Species 	&	DM	&	MM	&	AN	& Structure &	Dipole Ref	\\
			&	(d)	&	(g/mol)	&		&		\\
			\hline\hline
			\rule{0pt}{3ex}AlCl	&	0.2	&	51.5	&	30	&	Diatomic	&	\cite{77RoMe.AlCl}	\\
			AlH	&	0.01	&	28.0	&	14	&	Diatomic	&		\\
			AsH$_3$	&	0.2	&	78.0	&	36	&	Non-linear	&	\cite{CCCBDBWebsite} 	\\
			BeH	&	0.2	&	10.0	&	5	&	Diatomic	&	\cite{68ChDa.BeH}	\\
			CH	&	0.4	&	13.0	&	7	&	Diatomic	&		\\
			CN	&	1.5	&	26.0	&	13	&	Diatomic	&	\cite{CCCBDBWebsite} 	\\
			NO 	&	0.2	&	30.0	&	15	&	Diatomic	&	\cite{CCCBDBWebsite}	\\
			SiH$_2$ & 0.1 & 30.1 &   16 &  Non-linear & \\
			\hline\hline
		\end{tabular}
	\end{table}

	\begin{table}[H]
	\caption{Molecular properties for the species with computed opacities presented in this work, where insufficient broadening parameters were found in the literature, and so those of OCS (see Table~\ref{t:sources_2}) were used instead. DM is the dipole moment, MM is the molar mass, and AN is the atomic number.}
	\label{t:ocs} 
	\centering  
	\begin{tabular}{llllll}
		\hline\hline
		\hline
		\rule{0pt}{3ex}Species 	&	DM	&	MM	&	AN	& Structure &	Dipole Ref	\\
		&	(d)	&	(g/mol)	&		&		\\
		\hline\hline
		\rule{0pt}{3ex}H$_2$S	&	1	&	34.1	&	18	&	Polar	&		\\
		N$_2$O	&	0.2	&	44.0	&	22	&	Linear	&		\\
		NO$_2$ 	&	0.3	&	46.0	&	23	&	Non-linear	&	\cite{CCCBDBWebsite} 	\\
		\hline\hline
	\end{tabular}
\end{table}

	\begin{table}[H]
		\caption{Molecular properties for the species with computed opacities presented in this work, where insufficient broadening parameters were found in the literature, and so those of PH$_3$ (see Table~\ref{t:sources_2}) were used instead. DM is the dipole moment, MM is the molar mass, and AN is the atomic number.}
		\label{t:ph3} 
		\centering  
		\begin{tabular}{llllll}
			\hline\hline
			\hline
			\rule{0pt}{3ex}Species 	&	DM	&	MM	&	AN	& Structure &	Dipole Ref	\\
			&	(d)	&	(g/mol)	&		&		\\
			\hline\hline
			\rule{0pt}{3ex}O$_3$ 	&	0.5	&	48.0	&	18	&	Non-linear	&		\\
			PH	&	$\sim$0.5	&	32.0	&	16	&	Diatomic	&	\cite{CCCBDBWebsite} 	\\
			\hline\hline
		\end{tabular}
	\end{table}

	\begin{table}[H]
		\caption{Molecular properties for the species with computed opacities presented in this work, where insufficient broadening parameters were found in the literature, and so those of HCl (see Table~\ref{t:sources_2}) were used instead. DM is the dipole moment, MM is the molar mass, and AN is the atomic number.}
		\label{t:hcl} 
		\centering  
		\begin{tabular}{llllll}
			\hline\hline
			\hline
			\rule{0pt}{3ex}Species 	&	DM	&	MM	&	AN	& Structure &	Dipole Ref	\\
			&	(d)	&	(g/mol)	&		&		\\
			\hline\hline
			\rule{0pt}{3ex}AlF	&	1.5	&	46.0	&	22	&	Diatomic	&	\cite{CCCBDBWebsite} 	\\
			CaF	&	1.2	&	59.1	&	29	&	Diatomic	&	\cite{01RaJeJu.CaCl}	\\
			HBr	&	0.8	&	80.9	&	36	&	Diatomic	&		\\
			HeH$^{+}$	&	$\sim$1.3	&	5.0	&	3	&	Diatomic	&	\cite{CCCBDBWebsite}	\\
			HI	&	0.4	&	127.9	&	54	&	Diatomic	&		\\
			MgH 	&	1.2	&	25.3	&	13	&	Diatomic	&	\cite{CCCBDBWebsite}	\\
			O$_2$	&	0	&	36.0	&	16	&	Diatomic	&		\\
			PS	&	0.6	&	32.0	&	31	&	Diatomic	&	\cite{92KaGr.PS}	\\
			SiH	&	1.2	&	29.1	&	15	&	Diatomic	&		\\
			\hline\hline
		\end{tabular}
	\end{table}
	
	\begin{table}[H]
		\caption{Molecular properties for the species with computed opacities presented in this work, where insufficient broadening parameters were found in the literature, and so those of NH$_3$ (see Table~\ref{t:sources_2}) were used instead. DM is the dipole moment, MM is the molar mass, and AN is the atomic number.}
		\label{t:nh3} 
		\centering  
		\begin{tabular}{llllll}
			\hline\hline
			\hline
			\rule{0pt}{3ex}Species 	&	DM	&	MM	&	AN	& Structure &	Dipole Ref	\\
			&	(d)	&	(g/mol)	&		&		\\
			\hline\hline
			\rule{0pt}{3ex}H$_3$O$^{+}$	&	$\sim$1.5	&	19.0	&	11	&	Non-linear	&	\cite{CCCBDBWebsite}	\\
			P$_2$H$_2$ (cis)	&	1.4	&	64.0	&	32	&	Non-linear	&		\\
			\hline\hline
		\end{tabular}
	\end{table}
	
	\newpage
	\begin{table}[H]
		\caption{Molecular properties for the species with computed opacities presented in this work, where insufficient broadening parameters were found in the literature, and so those of HF (see Table~\ref{t:sources_2}) were used instead. DM is the dipole moment, MM is the molar mass, and AN is the atomic number.}
		\label{t:hf} 
		\centering  
		\begin{tabular}{llllll}
			\hline\hline
			\hline
			\rule{0pt}{3ex}Species 	&	DM	&	MM	&	AN	& Structure &	Dipole Ref	\\
			&	(d)	&	(g/mol)	&		&		\\
			\hline\hline
			\rule{0pt}{3ex}CP		&	2.1	&	43.0	&	21	&	Diatomic	&		\\
			CS		&	2.0	&	44.1	&	22	&	Diatomic	&	\cite{CCCBDBWebsite} 	\\
			FeH		&	2.0	&	56.9	&	27	&	Diatomic	&	\cite{07ChStMe.CrH}	\\
			LiH$^{+}$	&	$\sim$2	&	8.0	&	4	&	Diatomic	&	\cite{CCCBDBWebsite}	\\
			MgF		&	1.8	&	43.3	&	21	&	Diatomic	&		\\
			NH	&	0.5	&	15.0	&	8	&	Diatomic	&	\cite{CCCBDBWebsite}	\\
			NS		&	1.8	&	46.1	&	23	&	Diatomic	&	\cite{CCCBDBWebsite} 	\\
			OH		&	1.7	&	17.0	&	9	&	Diatomic	&	\cite{CCCBDBWebsite} 	\\
			OH$^{+}$	&	$\sim$2	&	17.0	&	9	&	Diatomic	&	\cite{CCCBDBWebsite}	\\
			PO	&	1.9	&	47.0	&	23	&	Diatomic	&	\cite{CCCBDBWebsite} 	\\
			ScH	&	1.7	&	46.0	&	22	&	Diatomic	&	\cite{CCCBDBWebsite} 	\\
			SiS		&	1.7	&	60.2	&	30	&	Diatomic	& \cite{CCCBDBWebsite} 	\\
			TiH		&	$\sim$2 	&	48.9	&	23	&	Diatomic	&	\cite{CCCBDBWebsite}	\\
			\hline\hline
		\end{tabular}
	\end{table}

	\begin{table}[H]
		\caption{Molecular properties for the species with computed opacities presented in this work, where insufficient broadening parameters were found in the literature, and so those of H$_2$CO (see Table~\ref{t:sources_2}) were used instead. DM is the dipole moment, MM is the molar mass, and AN is the atomic number.}
		\label{t:h2co} 
		\centering  
		\begin{tabular}{llllll}
			\hline\hline
			\hline
			\rule{0pt}{3ex}Species 	&	DM	&	MM	&	AN	& Structure &	Dipole Ref	\\
			&	(d)	&	(g/mol)	&		&		\\
			\hline\hline
			\rule{0pt}{3ex}CH$_3$Cl	&	1.9	&	50.5	&	26	&	Non-linear	&		\\
			CH$_3$F	&	1.9	&	34.0	&	18	&	Non-linear	&	\cite{CCCBDBWebsite}	\\
			H$_2$O$_2$	&	2.3	&	34.0	&	18	&	Non-linear	&		\\
			HNO$_3$	&	2.2	&	63.0	&	32	&	Non-linear	&	\cite{CCCBDBWebsite}	\\
			\hline\hline
		\end{tabular}
	\end{table}

	\begin{table}[H]
		\caption{Molecular properties for the species with computed opacities presented in this work, where insufficient broadening parameters were found in the literature, and so those of HCN (see Table~\ref{t:sources_2}) were used instead. DM is the dipole moment, MM is the molar mass, and AN is the atomic number.}
		\label{t:hcn} 
		\centering  
		\begin{tabular}{llllll}
			\hline\hline
			\hline
			\rule{0pt}{3ex}Species 	&	DM	&	MM	&	AN	& Structure &	Dipole Ref	\\
			&	(d)	&	(g/mol)	&		&		\\
			\hline\hline
			\rule{0pt}{3ex}AlO	&	4.2	&	43.0	&	21	&	Diatomic & \cite{CCCBDBWebsite}	\\
			CaH	&	2.9	&	41.1	&	21	&	Diatomic	&	\cite{06HoUr.CaH}	\\
			CaO	&	8.7	&	56.1	&	28	&	Diatomic	&	\cite{14YuTr.CaO}	\\
			CrH	&	3.9	&	53.0	&	25	&	Diatomic	&	\cite{CCCBDBWebsite}	\\
			KCl	&	10.2	&	74.6	&	36	&	Diatomic	&	\cite{CCCBDBWebsite}	\\
			KF	&	8.6	&	58.1	&	28	&	Diatomic	&	\cite{CCCBDBWebsite} 	\\
			LiCl	&	7.1	&	42.4	&	20	&	Diatomic	&	\cite{CCCBDBWebsite} 	\\
			LiF	&	6.3	&	25.9	&	12	&	Diatomic	&	\cite{CCCBDBWebsite} 	\\
			LiH	&	5.9	&	8.0	&	4	&	Diatomic	&	\cite{CCCBDBWebsite} 	\\
			MgO	&	6.2	&	40.3	&	20	&	Diatomic	& \cite{CCCBDBWebsite}	\\
			NaCl	&	9.0	&	58.4	&	28	&	Diatomic	&	\cite{CCCBDBWebsite} 	\\
			NaF	&	8.1	&	42.0	&	20	&	Diatomic	&	\cite{CCCBDBWebsite} 	\\
			NaH	&	6.0	&	24.0	&	12	&	Diatomic	&	\cite{CCCBDBWebsite} 	\\
			PN	&	2.8	&	45.0	&	22	&	Diatomic	&	\cite{CCCBDBWebsite} 	\\
			SH	&	2.7	&	33.1	&	17	&	Diatomic	&		\\
			SiO	&	3.1	&	44.1	&	22	&	Diatomic	&	\cite{13BaElKo.SiO}	\\
			TiO	&	$\sim$3	&	63.9	&	30	&	Diatomic	&	\cite{90BaLaKo.TiO}	\\
			VO	&	3.4	&	66.9	&	31	&	Diatomic	&	\cite{91SuFrLo.VO}	\\
			\hline\hline
		\end{tabular}
	\end{table}


\end{appendix}

\end{document}